\input harvmac
\input epsf
\noblackbox
%
\def\ntp{n^\prime_T}
\def\zp#1{z^{\prime#1}}
\def\sx#1{s^{n_#1}}
\def\tv{\tilde{v}}
\def\sut#1{\tilde{SU}_2^{(#1)}}
\def\mus{\mu^\star}
\def\ss{\scriptstyle}
\def\codim{{\rm codim}}
\def\zt{\tilde{z}}
\def\tIF{{\bf \tilde{F}}}
\def\la{\lambda}\def\zh{\hat{z}}
\def\lra{\leftrightarrow}
\def\chull#1{{\rm convex \ hull\ }\{ #1 \}}
\def\dmn{\Delta_{M_{n+1}}}\def\dwn{\Delta^*_{W_{n+1}}}
\def\dzn{\Delta^*_{Z_{n}}}
\def\nus{\nu^\star}\def\nupr{\nu^\prime}

\def\wc{{\cal W}}\def\pwc{{p_{\wc}}}

\def\hx#1{{\hat{#1}}}

\def\IW{{\bf WP}}\def\IFO{{\bf F}}
\def\Ds{\Delta^\star}
\def\abstract#1{
\vskip .5in\vfil\centerline
{\bf Abstract}\penalty1000
{{\smallskip\ifx\answ\bigans\leftskip 2pc \rightskip 2pc
\else\leftskip 5pc \rightskip 5pc\fi
\noindent\abstractfont \baselineskip=12pt
{#1} \smallskip}}
\penalty-1000}
\def\us#1{\underline{#1}}
\def\hth/#1#2#3#4#5#6#7{{\tt hep-th/#1#2#3#4#5#6#7}}
\def\nup#1({Nucl.\ Phys.\ $\us {B#1}$\ (}
\def\plt#1({Phys.\ Lett.\ $\us  {B#1}$\ (}
\def\cmp#1({Comm.\ Math.\ Phys.\ $\us  {#1}$\ (}
\def\prp#1({Phys.\ Rep.\ $\us  {#1}$\ (}
\def\prl#1({Phys.\ Rev.\ Lett.\ $\us  {#1}$\ (}
\def\prv#1({Phys.\ Rev.\ $\us  {#1}$\ (}
\def\mpl#1({Mod.\ Phys.\ Let.\ $\us  {A#1}$\ (}
\def\ijmp#1({Int.\ J.\ Mod.\ Phys.\ $\us{A#1}$\ (}
\def\br{\hfill\break}\def\ni{\noindent}
\def\cx#1{{\cal #1}}\def\al{\alpha}\def\IP{{\bf P}}
\def\tx#1{{\tilde{#1}}}\def\bx#1{{\bf #1}}
\def\ov#1#2{{#1 \over #2}}
\def\al{\alpha}
\def\subsubsec#1{\ \br \noindent {\it #1} \br}
\def\eps{\epsilon}
\lref\VW{C. Vafa and E. Witten, Nucl. Phys. Proc. Suppl. {$\us {46}$} 
(1996) 225.} 
\lref\KMV{S. Katz, P. Mayr and C. Vafa,
Adv. Theor. Math. Phys. $\us {1}$ (1998) 53.}
\lref\SDS{A. Klemm, 
W. Lerche, P. Mayr, C. Vafa, N. Warner,
                \nup 477 (1996) 746.}
\lref\KKV{S. Katz, A. Klemm and C. Vafa,
\nup 497 (1997) 173.}

\Title{\vbox{
\rightline{\vbox{\baselineskip12pt
\hbox{CERN-TH/98-365}
\hbox{NSF-ITP-98-099}
\hbox{hep-th/9811217}}}}}
{Heterotic String/F-theory Duality } 
\vskip-1cm\centerline{{\titlefont from Mirror Symmetry}}\vskip 0.3cm
\centerline{P. Berglund\foot{berglund@itp.ucsb.edu} and  
P. Mayr\foot{Peter.Mayr@cern.ch}}
\vskip 0.6cm
\centerline{$^1$ \it Institute for Theoretical Physics, 
University of California, Santa Barbara, CA 93106, USA}
\vskip 0.0cm
\centerline{$^2$ \it Theory Division, CERN, 1211 Geneva 23, 
Switzerland}
\vskip -0.8cm
\abstract{\ni
{}We use local mirror symmetry in type IIA string compactifications
on Calabi--Yau $n+1$ folds $X_{n+1}$ to construct vector bundles 
on (possibly singular) elliptically fibered Calabi--Yau $n$-folds 
$Z_n$. The 
interpretation of these data as valid classical solutions  of the
heterotic string compactified on $Z_n$ proves F-theory/heterotic
duality at the classical level. Toric geometry is used to establish
a systematic dictionary that assigns to each given toric $n+1$-fold $X_{n+1}$
a toric $n$ fold $Z_n$ together with a specific family of sheaves on it.
This allows for a systematic construction of phenomenologically
interesting $d=4$ $N=1$ heterotic vacua, e.g. 
on deformations of the tangent bundle, with 
grand unified and $SU(3)\times SU(2)$ gauge groups.
As another application we find
non-perturbative gauge enhancements of the heterotic string on singular 
Calabi--Yau manifolds and new non-perturbative dualities relating
heterotic compactifications on different manifolds. }
\Date{\vbox{\hbox{\sl {November 1998}}
}}
\goodbreak

\parskip=4pt plus 15pt minus 1pt
\baselineskip=15pt plus 2pt minus 1pt

\newsec{Introduction}
The heterotic string compactified on a Calabi--Yau three-fold $X_3$
has been the phenomenologically most promising candidate amongst 
perturbatively defined string theories for quite some 
time~\ref\candelas{P. Candelas,
G. Horowitz, A. Strominger and E. Witten, \nup 258 (1985) 46.}. In
particular, compactifications with $(0,2)$ supersymmetry can easily lead
to  realistic gauge groups \ref\wito{E. Witten, \nup 268 (1986) 79.}. The
definition of the theory involves the understanding  of a 
suitable stable vector bundle $V$ on $X_3$, which turns out to be
a very difficult problem, however.

A new promising approach to the problem is given by the
duality to F-theory 
\ref\vafaf{C. Vafa, \nup 469 (1996) 403. }.
The basic duality is  between F-theory on elliptically fibered K3 and the
heterotic string on $T^2$ in eight dimensions\lref\Sen{A. Sen, 
\nup 475 (1996) 562.}~\refs{\vafaf,\Sen}. 
Lower dimensional dualities are obtained
by "fibering the eight-dimensional duality" with the result that 
F-theory on an elliptically and K3 fibered $n$+1-dimensional Calabi--Yau
manifold $X_{n+1}$ is dual to the heterotic string on an elliptically
fibered $n$-dimensional Calabi--Yau $Z_n$~\ref\MVi{C. Vafa and D. Morrison, 
\nup 473 (1996) 74;\nup 476 (1996) 437.}.

Quantum corrections of various kinds on both sides are expected 
to challenge the usefulness of these dualities for the case
of minimal supersymmetry. However, to not put the cart before the
horse, the first task should be to determine dual compactifications
in a classical sense, by specifying to each Calabi--Yau manifold
$X_{n+1}$ for the F-theory compactification the dual manifold 
$Z_n$ together with an appropriate bundle\foot{Or more generally,
reflexive or coherent sheaves 
\ref\DGM{J. Distler, B.R. Greene and D.R. Morrison, \nup 481 (1996) 289.}
.} $V$ on it. A big step in this direction 
has been done in the mathematical analysis of 
\ref\Loo{E. Looijenga, Invent. Math. {$\us {38}$} (1977) 17;
Invent. Math. {$\us {61}$} (1980) 1.}%
\ref\FMW{R. Friedman, J.W. Morgan and E. Witten, \cmp 187 (1997) 679.}%
\ref\FMWii{R. Friedman, J.W. Morgan and E. Witten,
{\it Principal G-bundles over elliptic curves}, alg-geom/9707004;
{\it Vector Bundles over Elliptic Fibrations}, alg-geom/9709029.}%
\ref\BJPS{M. Bershadsky, A. Johansen, T. Pantev and V. Sadov,
\nup 505 (1997) 165.},
where the moduli space of holomorphic $H$ bundles on elliptically
fibered manifolds has been determined.

In this paper we take a rather different route to F-theory/heterotic
duality. We define vector bundles on general elliptically fibered 
Calabi--Yau $n$-folds purely in a type IIA language. Specifically,
the type IIA theory compactified on an elliptically and K3 fibered 
Calabi--Yau manifold $X_{n+1}$ describes holomorphic $H$ bundles on an
elliptically fibered Calabi--Yau manifold $Z_n$ on general grounds,
without any reference to a heterotic dual. As the starting point 
consider a type IIA compactification on K3$\times
T^2$ where the K3 is elliptically fibered and has a singularity of type $H$.
Part of the moduli space $\cx M_{IIA}$ is identified with the moduli space 
$\cx M_{T^2}$ of Wilson lines on $T^2$.
The $R$-symmetry of the $N=4$ supersymmetry of this compactification 
provides identifications in $\cx M_{IIA}$ which in particular 
relate K\"ahler deformations of the singularity $H$ in the elliptic fibration
of the  K3 to $\cx M_{T^2}$ 
\KMV. The correspondence can
be made precise by considering a certain local limit. Application 
of local mirror symmetry maps this description of $\cx M_{T^2}$ in terms
of K\"ahler moduli to a description of $\cx M_{T^2}$ in terms of 
complex deformations of a local mirror geometry\foot{Here 
and in the following a subscript denotes the complex
dimension of a geometry.} $\cx W_2$.
In particular,
$\cx W_2$ gives a concrete description of the elliptic curve $T^2$ and
a flat $H$ bundle on it, where $H$ is the type of the original singularity
we started with. Combining this construction with an adiabatic
argument as in
\VW,  we can use an equivalent limit of a type IIA compactification on a
Calabi--Yau
$n+1$-fold $X_{n+1}$ rather than K3, 
to describe deformations of a weighted projective 
bundle $\tx {\cx W}\to Z_n$ on the elliptically 
fibered manifold $Z_n$. These data
define a family of vector bundles $V$ on $Z_n$ \FMW.

Taking the small fiber limit of $X_{n+1}$, which does not interfer
with the limit extracting the submoduli space of
the holomorphic  bundle on $Z_n$, we arrive at a similar conclusion for 
the  F-theory vacua in two dimensions higher. F-theory/heterotic
duality reduces to the mere statement that we can now interprete
the stable $H$ bundle on $Z_n$ as a classical solution of the
heterotic string. 

Our approach improves in various aspects the previous understanding
and use of heterotic/F-theory duality. First note that rather
than comparing properties of two supposedly dual theories such 
as the topological data of line bundles in \FMW, we derive
F-theory/heterotic duality from known, classical physics of 
the type IIA compactification.
On the practical side, the method allows us to engineer a bundle
with any given structure group $H$ on any elliptically fibered
Calabi--Yau manifold $Z_n$ with no restriction on the smoothness of
the elliptic fibration. Using the powerful concept of toric geometry 
we can give a systematic construction of how to build the Calabi--Yau
manifold  $X_{n+1}$ from a few elementary building blocks. Subtleties
arising from singularities in the elliptic fibrations are 
taken care of by the toric framework. 

That a purely classical type IIA framework can be used to 
provide a geometric construction of a supersymmetric quantum theory
has a well-known prehistory. In the geometric realization of the
result of Seiberg and Witten 
\ref\SW{N. Seiberg and  E. Witten, \nup 426 (1994) 19, 
erratum: ibid {$\us {430}$} (1994) 396;
\nup 431 (1994) 484.}
on $N=2$ supersymmetric Super-Yang--Mills (SYM)
theory, the moduli space is
described 
in terms of periods of 
a Riemann surface $\Sigma$. The a priori surprising
appearance of the complex geometry $\Sigma$ is explained by the fact
that it appears as the mirror geometry of a type II  
Calabi--Yau three-fold 
compactification \ref\KKLMV{S. Kachru, A. Klemm, W. Lerche, 
P. Mayr and C. Vafa,
                \nup 459 (1996) 537.}\SDS\KKV\KMV.
Once this relation is recognized,
it is much simpler and much more general to obtain the  exact answer
for SYM theories in the type II setup, with the result that for general
gauge group $H$, the Seiberg--Witten geometry is a Calabi--Yau
threefold rather than a curve \KMV, as anticipated by the string
point of view.
In the present case, the mathematical analysis of \FMW\ leads to 
a formulation of the moduli space of $H$ bundles on an elliptically
fibered Calabi--Yau $Z_n$  in terms of deformations of 
seemingly unrelated geometries of various dimensions for various $H$ \FMW.
Again these geometries are identified in the present paper
directly with a physical type IIA 
compactification. These geometries are well-understood and together
with toric geometry provide an easy kit to construct any given 
combination of a family of bundles and a manifold. 
We will formulate our construction in terms of Calabi--Yau manifolds represented as toric
hypersurfaces, which also in many other respects is the most useful
and most general representation of these manifolds.

The organization of this paper is as follows. In sect. 2 we 
explain the basic set up, in particular the local mirror map, and give
an outline of the general recipe for lower-dimensional theories.
In sect. 3 we define the toric geometries and their mirrors 
for the eight-dimensional 
case and describe how the polyhedron
$\Delta^*_{X_n+1}$ associated to the toric manifold
$X_{n+1}$  in the description of toric geometry, encodes the toric 
manifold $Z_n$ in terms of a projection.
In sect. 4 we complete the eight-dimensional
dictionary between toric K3 manifolds and $H$ bundles on an elliptic 
curve.
In sect. 5 we illustrate the six-dimensional case, giving a
dictionary between local degeneration of Calabi--Yau three-folds $X_3$
on the one side and a family of vector bundles with structure group $H$
on a K3 manifold $Z_2$. We find that a singular heterotic manifold
can lead to non-perturbative symmetry enhancement for appropriate 
choice of gauge bundle. 
Many of these theories turn out to have a non-perturbative
dual with a gauge bundle with different structure group on a 
different K3 manifold. 
In sect. 6 we describe vector bundles on Calabi--Yau three-folds
which may serve as a classical vacuum of a four-dimensional $N=1$ 
heterotic string. We discuss 
compactification on the tangent bundle generally in $10-2n$ dimensions 
and find similar non-perturbative equivalences as those in the
six dimensions. Some phenomenological interesting configurations with 
gauge groups $E_8,\; E_7,\; E_6,\; SO(10),\; SU(5)$ and 
$SU(3)\times SU(2)$ are considered in sect. 7. We end with our 
conclusions in sect. 8.

\newsec{Holomorphic bundles on elliptically fibered Calabi--Yau's 
from mirror symmetry}
For the class of heterotic vacua with F-theory duals we can
restrict to the case of vector bundles on elliptically
fibered manifolds. For smooth fibrations,
the vector bundle $V$ on $Z_n$ can then
be thought of in a fiberwise way as the situation where the
data of a flat 
$H$ bundle on the elliptic curve $E$ varies over the points on the base.
This case has been analyzed mathematically 
in \FMW\BJPS. We then analyze the type IIA
aspect, stressing the natural appearance of the mirror manifold, $W_2$, and
how the moduli space of $H$ bundles is described in terms of complex
structure deformations of $W_2$. From this we are lead to the local
mirror limit and how the construction generalizes to lower dimensional
theories.

\subsubsec{The heterotic perspective}
Let us recall the basic construction of ref.\FMW. In \Loo\
it was shown that the moduli space $\cx M_E$ of a holomorphic $H$ bundle 
on the elliptic curve $E$ is  a weighted projective space 
$W=\IW^r_{s_0,\dots,s_{r}}$, where $r$ is the rank of $H$ and 
$s_i$ are the Dynkin numbers of the Dynkin diagram of the 
twisted Kac-Moody algebra dual to $H$. In \FMW, the moduli 
space $\cx M_Z$ of $H$ bundles on elliptically fibered manifolds $Z$
(not necessarily Calabi--Yau) is described in terms of
families of elliptic curves $E_b$ together with a bundle 
$V_b$ on $E_b$ varying over the base $B$ of the elliptic fibration 
$\pi:Z\to B$. Here $b$ parametrizes the base $B$. 
Part of the data  of the bundle $V$ on $Z$ is described by a bundle
$\tx {\cx W}$  of weighted projective spaces over $B$. Restriction to a point
$b\in B$ gives a weighted projective space $W_b$ as above, together with an
elliptic curve $E_b$. The bundle $\tx{\cx W} $ is given by a projectivization of
the bundle $\Omega$ 
\eqn\rfmwi{
\Omega=\cx O \oplus(\oplus_{i=1}^r \cx L^{-d_i})\; ,}
where $\cx L^{-1}$ is the normal bundle of $B$ and the $d_i$
are the degrees of the independent Casimir operators of $H$.

For example, in the case $G=SU(n)$, the zero of a section of 
$\tx{\cx W}$ determines a "spectral cover" $C\subset Z$, a 
codimension one submanifold in $Z$. $C$ intersects
an elliptic fiber $E_b$ at $n$ points  which define 
a holomorphic $SU(n)$ bundle by identifying $E_b$ with its Jacobian.

The information provided by a section of $\tx{\cx W}$ fixes only
the part of the data of $V$ which determines the restriction 
$V|_{E_b}$ to fibers of $\pi:Z\to B$. The information about
the non-trivial twisting is contained in a line bundle 
$\cx S$ on $C$ \FMW. 
If $C$ is non-simply connected, the
definition of $\cx S$ requires\foot{Additional
data related to singularities in the fiber product $Z\times_B Z$
are required if the complex dimension of $B$ is larger
than one. See
\ref\CD{G. Curio and R.Y. Donagi, \nup 518 (1998) 603.} for a partial
identification of these data and refs. \FMW\BJPS
\ref\CU{G. Curio, \plt 435 (1998) 39.} for comments on the
data related to additional singularities in the construction.} the specification of 
holonomies classified by the Jacobian of $C$. In the F-theory
compactification on $X$ 
this information corresponds to a point in the torus
$H^3(X,\bx R)/H^3(X,\bx Z)$. The moduli space $\cx M_Z$ of 
the $H$ bundles on $Z$ is thus fibered over the basis
$\cx Y$, the space of sections of $\tx{\cx W}$.   

In the duality between the heterotic string on $Z$ and F-theory
on $X$, the geometric data of $X$ fix only the bundle $\tx{\cx W}$.
This is clear upon further compactification on a two torus,
where the torus $H^3(X,\bx R)/H^3(X,\bx Z)$ gives rise to 
RR fields of the dual type IIA compactification, which are
non-geometric.

\subsubsec{The type IIA perspective}
The moduli space $\cx M_E$ of flat $H$ bundles on an elliptic curve $E$
appears in a different context in the type IIA compactification
on $K3\times T^2$, corresponding to the Wilson lines on $T^2$.
As observed in \KMV, this implies that complex deformations 
of K3, K\"ahler deformations of K3 and Wilson lines on the elliptic
curve $E$ can lead to equivalent moduli spaces. Specifically,
the three different kinds of deformations can be identified with
the three adjoint scalar fields in the $N=4$ vector multiplet
in four dimensions. Since there are elements of the $SO(6)$ R-symmetry which
rotate the three deformations into each other, one can infer
an equivalence of moduli spaces under certain conditions.
First note that if we choose a specific algebraic representation
$M_2$ for a K3, the K\"ahler deformations of $M_2$ are 
equivalent to complex deformations of the {\it mirror} manifold $W_2$,
so we may have to switch representations of K3, when applying
R-symmetry transformations. Similarly, the moduli space of $H$ Wilson
lines on $T^2$ will appear in a type IIA compactification on 
a K3 $Z_2$ with an $H$ singularity. The generic theory is mapped by 
the R-symmetry
to a type IIA compactification on a {\it smooth} K3, $M_2$ or $W_2$, in 
which an $H$ singularity is deformed in either K\"ahler or complex structure.

The moduli space $\cx M_E$ of $H$ Wilson lines on $T^2$ contains the geometric
deformations of $T^2$. In order that these moduli can be equivalent to
K\"ahler deformations of $M_2$, the latter has to be elliptically
fibered, with the two classes of the fiber and the base related
to the moduli of $T^2$. Similarly, to have an equivalent 
representation in terms of complex deformations of a K3, $W_2$,
we have to require the mirror $M_2$ to be elliptically fibered.

Moreover, in the full string moduli space these deformations are 
intertwined with each other, so we have to consider a special 
boundary in moduli space where the deformations are independent.
For K\"ahler deformations of a K3, $M_2$, this requires that we restrict
to a region in moduli space where the {\it local} deformations 
of a single $H$ singularity decouple from the rest of the global geometry.
The corresponding limit in the mirror manifold that gives
a representation of $\cx M_E$ as
{\it complex} deformations is given by translating the previous local limit in
K\"ahler moduli to a limit in complex moduli. It is defined by
the action of mirror symmetry on the moduli spaces of $M_2$ and
$W_2$. In the following this limit in $W_2$ will be denoted the local 
mirror limit  and the associated local geometry $\cx W_2$.

After having understood the relation between the moduli space of
elliptically fibered K3 and flat bundles on $E$, we can
discard the $T^2$ in the above discussion and consider the 
six-dimensional type IIA compactification on the K3, $M_2$. 
If both $M_2$ and $W_2$ are elliptically fibered, 
the complex and K\"ahler deformations of $M_2$ will
describe two vector bundles 
$V,V^\prime$ with structure groups $H,\; H^\prime$ on (different) tori $E,
\; E^\prime$, respectively. Because of the elliptic fibration we can 
moreover consider  the eight-dimensional small fiber limit of 
F-theory. The local mirror limit in complex structure
now describes $H$ Wilson lines on $T^2$.

Our strategy to describe an $H$ bundle on an elliptic curve $E$ 
is now as follows: we start with an elliptically fibered 
K3 manifold $M_2(H)$ with a singularity of type $H$ in
the elliptic fibration. Applying mirror symmetry to
$M_2(H)$ we obtain a mirror K3, $W_2(H)$, with the 
roles of complex and K\"ahler deformations exchanged.
In the local mirror limit we obtain a complex geometry $\wc_2$ whose 
deformations describe the $H$ bundle on $E_H$. Fibering the
local geometry $\wc_2$ over a complex base manifold $B$ we 
obtain an $H$ bundle on the elliptically fibered manifold $Z\to B$
with fibers $E_b$, $b\in B$.

\subsubsec{Type IIA/F-theory/heterotic duality}
Up to now we have obtained the moduli space of $H$ bundles on $T^2$ as
complex deformations of $\cx W_2$.
In addition we have K\"ahler deformations describing another
flat bundle on $T^2$.
This is also the classical moduli space of the heterotic string on 
$T^2\times T^2$ with a vector bundle that splits over the
two $T^2$ factors. In the small fiber limit this is reduced to
the heterotic string on $T^2$.
The statement that the full moduli spaces are equivalent is 
the conjectured duality between the heterotic string and the corresponding
type IIA/F-theory. In six dimensions, we can interpret the 
complex and K\"ahler deformations of the elliptically fibered
manifold $W_2$ as two sets of Wilson lines $W_I$ and $W_{II}$
with structure groups $H_I$ and $H_{II}$ 
on a $T^2_I\times T_{II}^2$ compactification of the heterotic
string, respectively. Specifically, the K3 $W_2$ provides a 
K\"ahler resolution of an $H_I$ singularity, whereas the 
complex deformations encode the resolution of an $H_{II}$ singularity.
The decompactification limit of
$T^2_{I}$ switches off the Wilson lines in ${H_I}$
and restores an $H_I$ gauge  symmetry.
It corresponds to the small fiber limit of $W_2$ which blows down
the K\"ahler resolution of the $H_I$ singularity in the elliptic
fibration. Specifically, the K\"ahler resolution replaces the singular elliptic
fiber by a collection of rk$(G)+1$ two-spheres which intersect according to the
affine Dynkin diagram of $H$. The classes of the blow-up spheres $C_i$
are related to the class $E$ of a generic fiber over a generic point by
\eqn\ebupo{
E=\sum_i s_i C_i\ ,
}
with $s_i$ the (positive) Dynkin indices. Thus blowing down the generic 
fiber blows down the spheres of the $H_I$ singularity leading 
to a gauge symmetry enhancement in the type IIA theory.

\subsec{The local mirror limit}
As mentioned before, the classical limit of the 
moduli space $\cx M_E(H)$ of $H$ bundles on $E$, as 
represented by a K\"ahler deformation of an $H$ singularity
in an elliptic fibration of a K3, $M_2$, corresponds to a limit
where we consider only the local deformations of the $H$
singularity and switch off the coupling to the global geometry.

However, we are really interested in taking the same limit 
for the {\it complex} deformations of the mirror $W_2(H)$. To be
specific consider the case of K3 manifolds dual to
the $E_8\times E_8$ string. In this case the K3 manifold
$M_2(H)$ has generically two singularities at the points $z=0$
and $z=\infty$ corresponding to the eight-dimensional gauge group 
$H=H_1\times H_2$ in the two $E_8$ factors.
We represent 
$M_2(H)$ and $W_2(H)$ as hypersurfaces in a toric variety.
The toric construction will be described in detail in the next section.
For the present discussion it is sufficient to know that 
the mirror manifold $W_2(H)$ is described as a hypersurface, 
given as the zero locus of a polynomial $p_{W_2(H)}$ in the
toric embedding space. We assert
that for the class of K3's dual to $E_8\times E_8$ string,
the polynomial $p_{W_2(H)}$ takes the general form 
\eqn\ktm{\eqalign{
p_{W_2(H)}&=p_0+p_{+}+p_{-}\ , \cr
p_0&=y^2+x^3+\zt^6+\mu xy\zt\ ,\cr
p_\pm &= \sum_{i=1}^{k} v^{\pm i} p_\pm^i \ ,
}}
where $y,x,\zt,v$ are specially chosen coordinates on the embedding space;
in particular $v$ is a coordinate on the base $\IP^1$.
Moreover $p_\pm^i$ are polynomials
in $y,x,\zt$ of homogeneous degree with respect to the scaling
action $(y,x,\zt)\to(\lambda^3 y,\lambda^2 x,\lambda \zt)$. Moreover
$\mu$ is a complex
parameter related to the complex structure of the 
elliptic curve $\hat{E}_H: p_0=0$.

The fact that $p_{W_2(H)}$ takes the above form is the way in which 
the complex geometry $\cx W_2(H)$ encodes the information that there
is a singularity in each of the two $E_8$ factors at the two points
$z=0$ and $z=\infty$, respectively. Note that in addition to taking the large
base limit in the original K3 $M_2(H)$ we have also to make a choice
of which of the two points $z=0$ or $z=\infty$ we want to concentrate
on. We claim that the local mirror corresponds to take a limit in 
the complex parameters such that
\eqn\lml{
p_-^i\to\eps^ip_-^i\ ,}
with $\eps\to 0$. The local mirror geometry $\wc$ is then given
by 
\eqn\ktmii{
p_{\wc}=p_0+p_+=0.}
The complex deformation of the geometry $\pwc=0$
describes the moduli space of a flat bundle $V$ 
compactified on the elliptic curve $E_H$ defined by 
the hypersurface $v=0$, $\hx E_H:\; p_0=0$. More
precisely $\hx E_H$ is the dual of the torus $E_H$;
in other words, we can think of $\hx E_H$ as the
Jacobian of $E_H$, which shares the complex structure
with $E_H$. The polynomial $p_+$ contains the information
about a bundle $V_+$ on $E_H$. E.g. for $H=SU(N)$, $N$ even, we will
have 
\eqn\sht{
p_+=v\;(\zt^N+\zt^{N-2}x+\zt^{N-3}y+\dots+x^{N/2}) \ .
}
As explained in more detail in a moment, we can integrate out
$v$ and obtain a geometry defined by the two equations 
$p_0=0\cap p_+=0$. This intersection gives $N$ points on $\hx E_H$,
which are interpreted as the values of the Wilson lines in the
Cartan algebra of $SU(N)$. These data specify uniquely the 
$SU(N)$ bundle $V_+$~\FMW.

Via duality, we will interpret this bundle as a bundle 
in the first $E_8$ factor of the heterotic string.
Since the original K3 had two singularities, 
the limit \lml\ must already include a choice of neighborhood. To describe
the neighbourhood of the second singularity, we simply rescale
the variable $v\to v\eps$ with the result that now the perturbations
in $p_+^i$ scale as $\eps^i$ while those in $p_-^i$ are constant.
The corresponding bundle $V_-$ can be interpreted as the bundle
in the second $E_8$ factor of the dual heterotic string.

The fact that the limit \lml\ is indeed the action on the
complex structure moduli obtained from action of mirror symmetry
on the local limit in the K\"ahler moduli space 
can be shown by a straightforward
analysis of the K\"ahler cone of $M_2(H)$ and the mirror map
between the K\"ahler moduli space of $M_2(H)$ and the complex
moduli of $W_2(H)$. This is described in Appendix A using the 
toric formulation introduced in the next section.

\subsec{Lower dimensional theories}
The above construction will be generalized to lower-dimensional 
dual pairs of F-theory on Calabi--Yau $W_{n+1}$ and the heterotic 
string on Calabi--Yau $Z_n$ by an application of the
adiabatic argument \VW. 
The geometry $W_2(H)$ describes an $H$ bundle 
over $E_H$ in the local limit. To obtain the description
of an $H$ bundle over an elliptically fibered Calabi--Yau $Z_n$
we can fiber the geometry $W_2(H)$ over an $n-1$ dimensional base $B_{n-1}$ to
obtain a Calabi--Yau $W_{n+1}$. In the local limit
we now get an $n$ dimensional geometry $\wc$ defined as in 
\ktm, but with the polynomials $p_\pm$ being functions
of the coordinates of the base $B_{n-1}$ (or rather sections of line bundles
on $B_{n-1}$). Similarly the bundle is now defined on the projection to $p_0=0$,
which  gives an $n$ dimensional 
Calabi--Yau $\hx Z_n$. We can identify $\hx Z_n$ with the 
dual\foot{With duality understood as the replacement of 
the elliptic fiber $E_H$ by the dual $\hx E_H$ parametrizing the 
Jacobian of $E_H$. For simplification we will drop the 
hat on $Z_n$ in the following.} $\hx Z_H$ of a 
dual heterotic compactification manifold $Z_H$. We will return to the
higher dimensional case later. For now we
note that the local limit is taken only in the fiber $W_2(H)$ 
(we choose to concentrate on the point with the singularity in 
the K3 fiber), but we retain the global structure of the elliptic
fibration over the base $B_{n-1}$. 

\newsec{Toric Construction}
\subsec{General remarks}
In the toric framework, a Calabi--Yau manifold $M_n$ is
described by an $n+1$ dimensional polyhedron $\Delta$.
The vertices $\nu_k$ of $\Delta$ lying on faces of codimension larger
than one correspond to divisors $x_k=0$ in $M_n$. 
Here $x_k$ is a coordinate on the toric ambient space associated
to $\nu_k$.
If $n=2$, that is $M_2$ is a K3 manifold, then a divisor is 
simply a holomorphic curve.
On the other hand, the K\"ahler resolution of a local $H$ singularity 
gives also rise to a collection of holomorphic spheres, intersecting 
according to the Dynkin diagram of $H$. Therefore the resolution
of $H$ corresponds to a set of vertices $\nu_k$ in the polyhedron 
$\Delta$ of $M_2(H)$. It is in this way that the construction of 
the K3 $M_2(H)$ and its mirror $W_2(H)$ can be phrased entirely in terms
of the polyhedra $\Delta$ and $\Delta^*$. For $n>2$, the divisors 
$x_k=0$ are no longer curves, but are dual to curves in $X_n$\foot{
For more details the reader is refered to the reviews and discussions 
of toric geometry in the physics literature
\ref\TRV{
R.S. Aspinwall, B.R. Greene and D.R. Morrison, \nup 416 (1994) 414;
P. Candelas, X. de la Ossa and S. Katz, \nup 450 (1996) 267;
S. Katz, P. Mayr and C. Vafa as quoted in \KMV;
N.C. Leung and C. Vafa, Adv. Theor. Math. Phys. $\us 2$ (1998) 91.
}.}.

In general the mirror manifold $W_n$, associated to $M_n$, can be obtained
using Batyrev's construction in terms of of the dual polyhedron 
$\Ds$
\ref\batms{V. Batyrev, Duke Math. Journ. $\us {69}$ (1993) 349.}.
The manifolds $M_n$ and $W_n$ are then defined by the zero of
a polynomial as:
\eqn\batpol{\eqalign{
p(M_n)&=p_\Delta=\sum_j a_j\; \prod_i x_i^{\langle \nu_i,\nus_j\rangle+1}\; ,\cr
p(W_n)&=p_\Ds=\sum_i b_i \; \prod_j x_j^{\langle \nu_i,\nus_j\rangle+1}\; ,}}
where the sum (product) runs over all vertices $\nu_i$ ($\nus_j$)
of $\Delta$ $(\Delta^*)$
which are on faces of codimension higher than one. The parameters
$a_j,\; b_i$ determine the complex structure.

If $N$ is the number of (relevant) 
vertices $\nu_i$ of $\Delta$, there will be 
$N-n-1$ non-trivial relations $\sum_i l_i^{(r)}\nu_i=0$ defined by the
vectors $l^{(r)}$. From \batpol\ it follows that the equation 
$p_\Delta$ is invariant under rescalings $x_i\to x_i\mu^{l_i^{(r)}}$.
Using these rescalings we can set 
$N-n-1$ coordinates to one in the appropriate patch and 
the remaining equation in $n+1$ variables defines a patch of the Calabi--Yau
manifold $M_{n}$ of dimension $n$. The number of such rescalings is
precisely the hodge number $h^{1,1}(M_{n})$~\foot{This assumes that
all the K\"ahler deformations are toric.}. This is of course 
a consequence of the fact that a vertex $\nu_i$ in $\Delta$ corresponds
to a divisor in $M_n$ which in turn defines a (1,1) form $K_r$ on $M_{n}$.
The same is true for $W_n$ with the roles of $\Delta$ and $\Ds$
interchanged.

{}From \batpol\ one can see that the relations $l^{(r)}$ translate
to relations 
\eqn\lmsr{
\prod_i y_i^{l^{(r)}_i}=1}
between the monomials 
$y_i=\prod_j x_j^{\langle \nu_i,\nus_j\rangle}+1$ of $p_\Ds$. In 
other words, given $p_\Delta$ and its scaling relations we can 
construct $p_\Ds$ by solving for these relations without having a 
dual polyhedron. In particular we can take the vertices associated
to the resolution of an elliptic singularity as in Table 1 and get an
equation for the mirror geometry, which describe a flat $H$ bundle
over an elliptic curve \KMV, for the reasons explained above. 

If these vertices are part of a polyhedron $\Delta$ of a 
compact Calabi--Yau manifold, we can alternatively use
Batyrev's construction \batpol\ to get the global mirror 
geometry. This global information contains in particular a
complete basis of divisors in $W_n$ and the associated rescalings
$l^{(r)*}$. The polynomial $p_\Ds$ obtained in this way agrees
with the one obtained from the local description in a certain
patch with some of the $x_i$ set to one. 

\subsec{Mirror geometries for elliptically fibered K3}
We proceed with a construction of $W_2(H)$ in terms of toric 
polyhedra 
and the derivation of the precise form of \ktmii\ for various $H$.

The toric resolution of singularities in an elliptic fibration has
been studied in \ref\Betal{M. Bershadsky et al., \nup 481 (1996) 215.}%
\ref\CF{P. Candelas and A. Font, \nup 511 (1998) 295.}\KMV. 
For the case $n=2$, we use the definitions 
{\ninepoint\eqn\verts{\eqalign{
&e_2=(0,-1,0),\; e_3=(0,0,-1), \cr
&f_1=(0,2,3),\; f_2=(0,1,2),\; f_3=(0,1,1)\cr
&v_0=(-1,2,3),\; 
v_k=(-1,2-[\ov{k+1}{2}],2-[\ov{k}{2}]),\quad
w_k=(-2,3-k,4-k),\; k>0\; ,\cr
&s_1=(-2,1,1),\;s_2=(-3,1,1),\quad t_1=(-3,1,2),\ t_2=(-4,1,2)\ ,\cr
&u_1=(-3,2,3),\; u_2=(-4,2,3),\; u_3=(-5,2,3),\; u_4=(-6,2,3)\; .
}}}

\ni
Moreover a tilde will denote a reversal of the sign of the first entry,
e.g. $\tilde{v}_0=(1,2,3)$. The toric data for an $H$ singularity 
in the elliptic fibration over a plane are \Betal:
\vskip .2cm
\vbox{\eqn\esr{
\vbox{\offinterlineskip\tabskip=0pt\halign{\strut
\vrule#&~~~$#$~~~\hfil&\vrule#
&~~~$#$~~~\hfil&\vrule#\cr
\noalign{\hrule}
&H 
&&\{\nu_i\}&\cr
\noalign{\hrule}
&SU(N)&&        v_1,\dots,v_{N-1}&\cr
&SO(7)&&        v_2,v_3,w_1&\cr
&SO(2N+5)&&        v_2,v_{N+1},w_1,\dots,w_N&\cr
&Sp(N)&&       v_1,v_3,\dots,v_{2N-1}&\cr
&SO(2N+6)&&    v_2,v_{N+1},v_{N+2},w_1,\dots,w_N&\cr
&G_2&&v_2,w_1&\cr
&F_4&& u_1,v_3,w_1,w_2&\cr
&E_6&&s_1,u_1,v_3,v_4,w_1,w_2&\cr
&E_7&&s_1,t_1,u_1,u_2,v_4,w_1,w_3&\cr
&E_8&&s_2,t_2,u_1,\dots,u_4,w_1,w_3&\cr
\noalign{\hrule}}}
}
\leftskip .5cm \rightskip .5cm 
\noindent{\ninepoint  \baselineskip=8pt  
{{\bf Table 1:} Vertices $\{\nu_i\}$ of the toric polyhedron
$\Delta_{local}=\chull{e_2,e_3,v_{0},\nu_i}$ for the resolution of an
$H$ singularity in the 
elliptic fibration over the plane.}}.}

\leftskip .0cm \rightskip .0cm \ni
Solutions of the relations \lmsr\ for many $H$ has been given in
\KMV\ and will be extended in the next section to all Lie groups. The generic
structure is as follows. Firstly we associate to the vertices $e_i,f_1$ 
of the elliptic fiber the monomials:
\eqn\msi{
e_2 \lra x^3,\ e_3 \lra y^2,\ f_1 \lra \zt^6.}
Moreover we associate to a vertex $\nu_i\in\Delta_{local}$ 
with first entry $\nu_{i,1}$ 
a monomial 
\eqn\vrel{v^{-\nu_{i,1}}f(y,x,\zt)\; ,} 
such that  $f(y,x,\zt)$ is a polynomial
which solves the obvious linear relations between the vertices. 
Note that $(y,x,\zt)$ will appear as homogeneous coordinates of the
elliptic curve $\hat{E}_H:p_0=0$ while $v$ defines a grading of the bundle
on $\hat{E}_H$.

Note that the (negative of) first entry of the vertices in \esr\ agrees
with the Dynkin label of the twisted Kac-Moody algebra 
associated to a simple Lie group $H$ shown in Fig. 1.
Thus the $v$ powers that appear are also in one to one correspondence
with the Dynkin labels. Moreover the number of monomials in 
$p_+^i$ is equal to the number of Dynkin labels equal to $i$.

\vskip 0.5cm
{\baselineskip=12pt \sl
\goodbreak\midinsert
\centerline{\epsfxsize 2.8truein\epsfbox{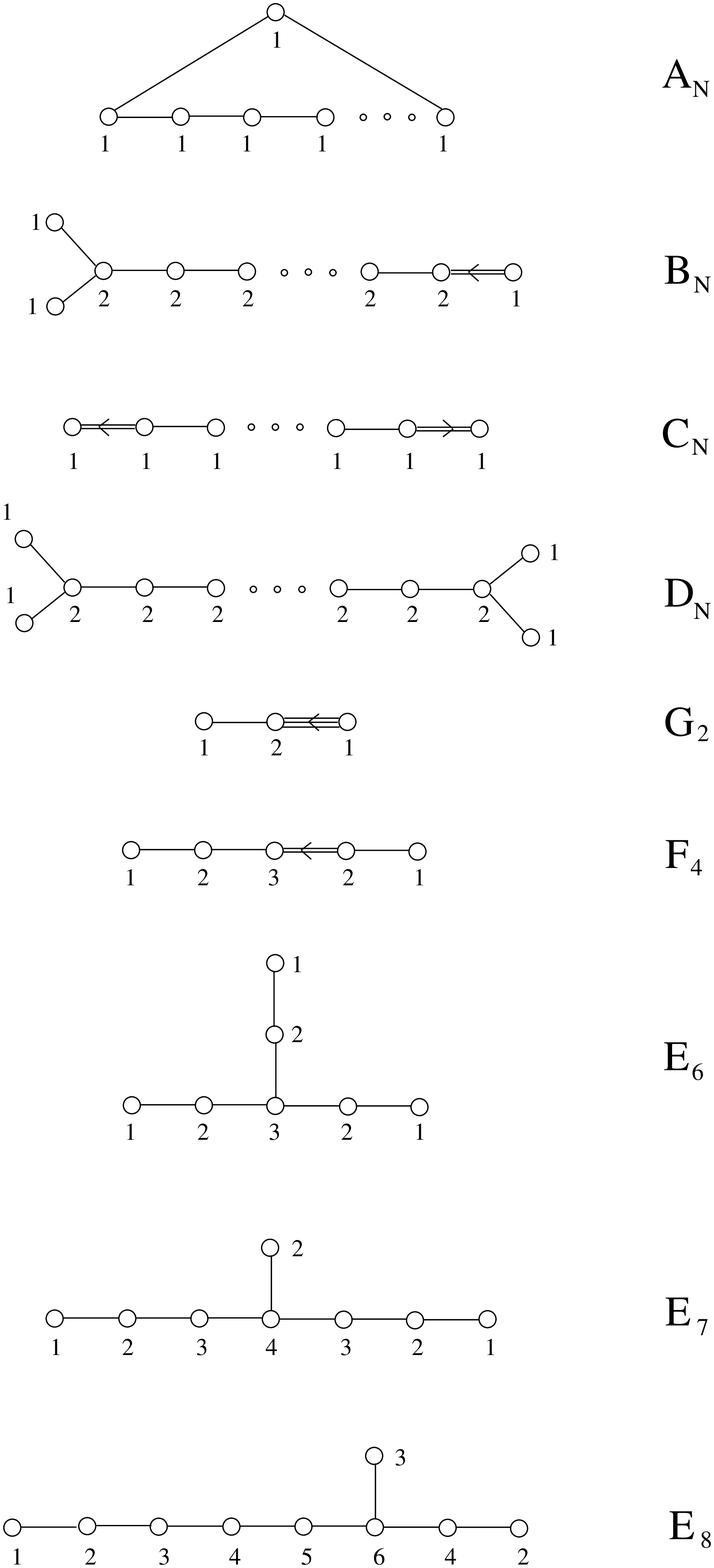}}
\leftskip 1pc\rightskip 1pc \vskip0.3cm
\noindent{\ninepoint  \baselineskip=8pt 
{{\bf Fig. 1:}
Dynkin diagrams for the duals of the untwisted Kac-Moody algebras. 
The integers denote the associated Dynkin labels for the affine root.}
}\endinsert}\ni

\subsubsec{Global K3 mirror manifolds}
While we can use the local description above to define any
flat $H$ bundle on the elliptic curve $\hat{E}_H$, viewing it
as a heterotic vacuum will of course imply restrictions on the possible
structure groups $H$. The point is that precisely if $H$ can be
embedded in a heterotic gauge group then the local geometry 
corresponding to 
$\Delta_{local}$ of Table 1 can be embedded into a 
global K3 geometry $M_2(H)$ corresponding to a larger polyhedron $\Delta$. 
In particular we can construct
a dual polyhedron $\Ds$ describing a global geometry $W_2$ that
contains the local mirror geometry in a patch. 
This will be very useful when
constructing fibrations of the local mirror geometry over
an $n-1$ complex dimensional base. 
If we construct a global K3 manifold $M_2$ with only an elliptic singularity
of type $H$ above a single point, we obtain the vertices of $\Ds$
shown in Table 2:\foot{We have performed a simple rotation of
basis in order to use the definitions \verts.}\vskip 0.5cm

\vbox{
\eqn\ktl{
\vbox{\offinterlineskip\tabskip=0pt\halign{\strut
\vrule#&\hfil~$#$~\hfil
&\vrule#& \hfil~$#$~\hfil&
&\vrule#& \hfil~$#$~\hfil&\vrule#&
$\;$\vrule#&\hfil~$#$~\hfil&\vrule#
& \hfil~$#$~\hfil&\vrule#
& \hfil~$#$~\hfil&\vrule#
\cr
\noalign{\hrule}
&H 
&&\{\nus_i\}
&&G_H&&&H 
&&\{\nus_i\}&&G_H
&\cr
\noalign{\hrule}
&SU(1)&&        u_4&&E_8&&&
SO(14)&&\tx v_0,        f_3,     v_3  &&\overline{SU(2)}&\cr
&SU(2)&&        u_2,w_3&&E_7&&&
SO(16)&&\tx w_1,       f_2,   v_3&&SU(1)&\cr
&SU(3)&&        s_1,u_1,v_4&&E_6&&&
SO(5)&&   w_1,w_3&&SO(11)&\cr
&SU(4)&&       v_4,w_1,w_2&&SO(10)&&&
SO(7)&&   w_1,w_2&&SO(9)&\cr
&SU(5)&&       v_0,v_2,v_3,v_4&&SU(5)&&&
SO(9)&&   v_3,w_1&&SO(7)&\cr
&SU(6)&&f_1,    v_1,v_3,v_4&&SU(2)\times SU(3)&&&
SO(11)&&   v_0,v_3&&SO(5)&\cr
&SU(7)&&\tx v_0,     f_3,     v_3,v_4&&SU(2)\times \overline{SU(2)}&&&
SO(13)&&f_1,  v_3 &&SU(2)&\cr
&SU(8)&&\tx w_1,         f_2,    v_3,v_4&&SU(2)&&&
SO(15)&&\tx v_0,         v_3 &&SU(1)&\cr
&SU(9)&&\tx u_1,\tx v_2,     v_3,v_4&&SO(14)\times\overline{SU(2)}&&&
G_2&&  u_1&&F_4&\cr
&Sp(3)&&      f_1,   w_3&&G_2\times SU(2)&&&
F_4&&   w_1&&G_2&\cr
&Sp(4)&&\tx w_1,           w_3&&SU(2)_2&&&
E_6&&v_0,v_1,v_2 &&SU(3)&\cr
&SO(10)&&   v_0,v_2,v_3&&SU(4)&&&
E_7&&v_0,v_1&&SU(2)&\cr
&SO(12)&&f_1,     v_1,v_3&&SU(2)\times SU(2)&&& 
E_8&&v_0&&SU(1)&\cr
\noalign{\hrule}}}
}
\noindent{\ninepoint  \baselineskip=8pt  
{{\bf Table 2:}
Vertices $\{\nus_i\}$ associated to 
the moduli space of $H$ bundles over $E_H$
in terms of complex geometries. The toric polyhedron is
$\Delta^*_H$=convex hull $\{\tx u_4,e_2,e_3 \} \cup \{ \nus_i \}$.
For $H=SU(9)$ $G_H$ is the commutant of $H$ in
$SO(32)$.
}}}

There is a nice property of
mirror symmetry when acting on the elliptically fibered K3 manifolds 
in Table 2 and their generalization with two singularities
at the two points $z=0$ and $z=\infty$ of the base $\IP^1$
(corresponding to an elliptic fibration with a $E_8\times E_8$
structure):
\vskip0.4cm

\leftskip 0.5cm\rightskip 0.5cm
\vbox{\ni\it $(*)$ Let $X_2(H_1,H_2)$ denote the elliptically fibered K3
manifold with singularities of type $H_1$ and
$H_2$ at $z=0$ and $z=\infty$, respectively. Then the  mirror of 
$X_2(H_1,H_2)$ is a K3 manifold of type $X_2(G_{H_1},G_{H_2})$, with
$G_{H_i}$ the commutant of $H_i$ in $E_8$.}

\leftskip 0cm\rightskip 0cm
\noindent
This result is not unexpected in view of the interpretation of 
K3 mirror symmetry in terms of orthogonal lattices 
\ref\KTMS{H. Pinkham, C. R. Acad. Sci. Paris $\us{284A}$ (1977) 615;\br
I.V. Dolgachev, J. Math. Sciences $\us{81}$ (1996) 2599.}.
It can be proven by a straightforward application of Batyrev's
construction of mirror manifolds in terms of dual polyhedra.
A similar statement applies to fibrations with only a single 
singularity corresponding to K3's dual to the $SO(32)$ heterotic
string. We have indicated the commutants $G$ in Table 2.
An over-lined $SU(2)$ denotes a special $SU(2)$ factor which appears
as a point in the hyperplane of the elliptic fiber; in this case there
is no decomposition as $H\times G$ in terms of maximal subgroups
(as e.g. in the breaking $E_8\to SU(7)$).

\subsec{Lower dimensional theories:\ Heterotic polyhedra 
from F-theory polyhedra}

To obtain a description of holomorphic bundles on elliptic (Calabi--Yau)
manifolds we need a toric description of a fibration of the local
mirror geometry $\cx W_2$ above a base manifold $B_{n-1}$. This is
a very simple process if $\cx W_2$ can be described in terms
of a polyhedron $\Delta^*$, which as we noted above is immediate
if the structure group $H$ fits into a heterotic gauge group.
In this case we have the embedding in a global K3 geometry 
$W_2$.
We will restrict to this simple case, which is also the physically
most interesting one in the following. The general case is 
more involved technically but can be treated very similarly.

In \ref\AKMS{A. C. Avram, M. Kreuzer,
M. Mandelberg and H. Skarke, \nup 494 (1997) 567.} it
was shown that a toric manifold $X_n$ defined by a polynomial as in
\batpol\ admits a fibration with Calabi--Yau fibers $Y_{k}$, if its polyhedron 
$\Delta_{X_n}$ contains the polyhedron of the fiber $Y_{k}$ as
a hypersurface of codimension $n-k$. Thus a fibration of $\cx W_2 \subset W_2(H)$
over an $n-1$ dimensional base $B_{n-1}$
is described by an $n+2$ dimensional polyhedron $\Delta^*_{W_{n+1}}$ that
contains $\Ds_{W_2}$ as a hyperplane. Specifically we can choose
coordinates such that the hypersurface $\{\nus_i\in\Delta^*:\
\nus_{i,j}=0,\ j=1,\dots,n-1\}$ contains the vertices $(0^{n-1},\nus_i(W_2))$,
with $\nus_i(W_2)$ the three-dimensional vertices described in the previous
section.

Since we have a well-defined global geometry we can get the
defining equation for the mirror manifold directly from \batpol,
rather than solving \lmsr. After a choice of variables -- 
corresponding to setting some of the toric variables $x_k$ to 
one or equivalently concentrating on the relevant local patch -- 
we obtain an expression precisely as in \ktm, but with the 
coefficients of the 
polynomials $p^i_\pm$ in $(y,x,\zt)$  
being functions of the toric coordinates
on $B_{n-1}$. In particular, $Z_n:p_0=0$ defines an $n$ dimensional
Calabi--Yau manifold. The holomorphic $H$ bundle is defined on $Z_n$
and we are free to interpret this data as a classical heterotic
vacuum.

We have used a limit of the $n+1$ dimensional 
toric geometries $W_{n+1}$ to describe a heterotic compactification
on an $n$ dimensional Calabi--Yau $Z_n$ with a prescribed vector bundle.
In toric terms, $W_{n+1}$ is given by an $n+2$ dimensional polyhedron
$\dwn$ and its dual $\dmn$, while the manifold $Z_n$ can be  described
by an $n+1$ dimensional polyhedron $\dzn$ and its dual. Let us see
how to get $\dzn$ directly from $\dwn$ by an appropriate projection. 

First recall that a hypersurface $\cx H$ in a polyhedron $\Delta$ corresponds
to a projection in the dual polyhedron $\Ds$. This is evident if
we choose coordinates where the $k$ dimensional hypersurface
$\cx H$ is described by 
vertices $\nu_i\in\Delta$ with the first $k$ entries equal to zero. The 
inner product $\langle \nu_i,\nu_j^*\rangle$ that determines the monomials
in \batpol\ does not depend on the first $k$ entries of the vertices
$\nu_j^*$, thus defining a projection in $\Delta^*$.

The "heterotic manifold" $Z_n$ is defined by $p_0=0$ which 
contains the monomials with zero power of $v$. Recall that 
in the K3 case the $v$ power is associated to the first entry $\nu_{i,1}$ of a
vertex in $\Delta$ and in the above conventions it will be the $n$-th
entry of the higher dimensional polyhedron $\dmn$, the dual 
polyhedron of $\dwn$. Thus {\it the heterotic manifold $Z_n$  
corresponds to a projection 
in the $n$-th direction of the polyhedron $\dwn$
and its mirror $Z^*_n$ to a hyperplane
$\nu_{i,n}$=0 of the polyhedron $\dmn$.}

\eqn\gpnot{\eqalign{
f:\; &W_{n+1}\to Z_n \cr
&\ p_0=\sum_{i} a_i \prod_j x_j^{\langle \nupr_i,\nus_j \rangle+1}\qquad 
\nupr_{i,n}=0  .}}

Note that the above implies 
that the mirror $M_{n+1}$ of the Calabi--Yau $W_{n+1}$ which 
is dual to the heterotic string on $Z_n$ admits a $Z^*_n$ fibration,
$\pi:M_{n+1}\to \IP^1$ with fibers $Z^*_n$, where $Z_n^*$ is the mirror
manifold of $Z_n$ !

\newsec{Local mirror limit of K3 manifolds}
{}From the toric construction of the elliptic singularities
over the plane, see sect. 3.2, and solving \lmsr\ using the
variables defined below eq.\msi, we can determine the 
polynomials $p_+^i$ appearing in the local limit \lml. 
Below we collect the results for the various choices of
structure group $H$ of the bundles.

The $H=SU(N)$ cases can be 
phrased in the general form \KMV,
\eqn\An{
p^1_+=a_1\zt ^N +a_2\zt ^{N-2}x +a_3\zt ^{N-3}y +\dots+
{a_Nx^{N/2}\brace
a_Nyx^{\ov{N-3}{2}}}\ ,}
where the $\{a_i\}$ are coordinates on the moduli space which is isomorphic
to $\IP^{N-1}$.
The geometry $p_e+v\; p^1_+=0$ 
describes a {\it two}-dimensional complex geometry. Note
that we have four coordinates, one equation, and one scaling relation
\eqn\sri{
(y,x,\zt ,v)\sim (\la^3 y,\la^2x,\la \zt ,\la^{6-N}v)\; .} This is 
different from the zero dimensional spectral cover description of the
moduli space of $SU(n)$ bundles obtained in \FMW\BJPS. However
as far as the complex structure moduli space is concerned, we
can integrate out linear variables, that is $v$ in the geometry
above, to obtain a zero dimensional geometry 
$$
p_E=0,\qquad p^1_+=0\ ,
$$
the spectral cover. Note that the situation is very similar 
to what happens in the case of moduli spaces of
$N=2$ $d=4$ SYM theories: the general complex geometry 
determining the exact solution is a Calabi--Yau three-fold \KMV\ 
but for the $SU(N)$ case it is natural to integrate out two
dimensions \KKLMV\SDS\ to get the Riemann surface of \SW.

For $SO(2N+1)$ the local limit can be phrased in the general
form
\eqn\Bn{\eqalign{
p^1_+&=b_1\zt ^{N-3}y+b_2\zt ^{N}+\eps\ b_3yx^{\ov{N+1}{2}-2}
+(1-\eps)\ b_3 x^{\ov{N}{2}}\ ,\cr
p^2_+&=a_1\zt ^{2N-6}+a_2\zt ^{2N-8}x+\dots+a_{N-2}x^{N-3}\ ,}}
with $\{b_1,b_2,b_3,a_1,\dots,a_{N-2}\}$ coordinates on the moduli space
$\IW^{N}_{1,1,1,2,\dots,2}$ of flat $B_{N}$ bundles on elliptic curve $E$.
Moreover $\eps=(0,1)$ for $N$ even (odd). This expression can be shown to be  
valid for any $N>4$ by using non-local Calabi--Yau two-fold geometries 
\ref\kmvii{S. Katz, P. Mayr and C. Vafa, unpublished}. The only other
case $SO(7)$ is slightly irregular and is obtained from
the $SO(9)$ case by discarding the term $a_2x$ in $p_+^2$. 

For $Sp(N)$ bundles we obtain
\eqn\Cn{
p^1_+=a_1\zt ^{2N} +a_2\zt ^{2N-2}x +a_3\zt ^{2N-4}x^2 +\dots+a_Nx^{N}
}
with $\{a_i\}$ coordinates on the moduli space $\IP^{N}$ of $C_N$ bundles.
The $Sp(N)$ case can be considered as a modding $y \to -y$ of the
$SU(2N)$ case in \An.

$SO(2N)$ bundles are described by a geometry $\cx W_2(D_N)$ with \KMV
\eqn\Dn{\eqalign{
p^1_+&=b_1\zt ^{N-3}y+b_2\zt ^{N}+c_1\zt ^{1-\eps}yx^{\ov{N+\eps}{2}-2}
+c_2 \zt ^\eps x^{\ov{N-\eps}{2}}\ ,\cr
p^2_+&=a_1\zt ^{2N-6}+a_2\zt ^{2N-8}x+\dots+a_{N-3}\zt ^2x^{N-4}\ .}}
Here $\{b_1,b_2,c_1,c_2,a_1,\dots,a_{N-3}\}$ are coordinates on the moduli space
$\IW^{N-1}_{1,1,1,1,2,\dots,2}$ of flat $D_{N-1}$ bundles on elliptic curve $E$,
and again $\eps=(0,1)$ for $N$ even (odd). 

For the exceptional group $G_2$ we find
\eqn\Gii{
p^1_+= a_1\zt ^3+a_2y\ ,\qquad p^2_+=b_1
}
with $\{a_1,a_2,b_1\}$ coordinates on the moduli space $\IW^2_{1,1,2}$ of
flat $G_2$ bundles. 

The geometry $\cx W_2(F_4)$ for the exceptional group $F_4$ is given by
\eqn\Fiv{
p^1_+= a_1\zt ^4+a_2x^2\ ,\qquad p^2_+=b_1\zt ^2+b_2x\ , \qquad p^3_+=c_1
}
where $\{a_1,a_2,b_1,b_2,c_1\}$ are coordinates on the moduli space 
$\IW^2_{1,1,2,2,3}$ of
flat $F_4$ bundles. 

Finally the geometries for the exceptional groups $E_n$ can be written as
follows\foot{There are several equivalent ways of parametrizing these
geometries, see \FMW\KMV.}. For $E_6$ we have
\eqn\Evi{
p^1_+=a_1\zt ^5+a_2\zt x^2+a_3xy\ ,\qquad p^2_+=b_1\zt ^4+b_2\zt y+b_3\zt ^2x\ ,
\qquad p^3_+=c_1\zt ^3\ ,}
parametrizing a $\IW^6_{1,1,1,2,2,2,3}$,\br
for $E_7$
\eqn\Evii{\eqalign{
p^1_+&=a_1\zt ^5+a_2xy\ ,\qquad p^2_+=b_1\zt ^4+b_2x^2+b_3y\zt\ ,\cr
p^3_+&=c_1\zt ^3+c_2x\zt\ , \qquad p^4_+=d_1\zt ^2 \ ,}}
defining a moduli space $\IW^7_{1,1,2,2,2,3,3,4}$ \br
and for $E_8$
\eqn\Eviii{\eqalign{
&p^1_+=a_1\zt ^5\ ,\qquad p^2_+=b_1\zt ^4+b_2x^2\ , \qquad 
p^3_+=c_1\zt ^3+c_2y\ , \cr
&p^4_+=d_1\zt ^2+d_2x \ ,\qquad
p^5_+=e_1\zt \ , \qquad p^6_+=f_1\ ,}}
giving a moduli space $\IW^8_{1,2,2,3,3,4,4,5,6}$. 
The $E_n$  geometries define two-dimensional  complex del Pezzo surfaces.

The parametrization of the complex geometries as above is
ambiguous in the sense that there are additional terms 
compatible with the scaling symmetries which can be 
absorbed by variable definitions. E.g. in the polynomial for
the elliptic curve $E:y^2+x^3+\zt ^6+yx\zt =0$ we can eliminate the
linear term in $y$ by a shift of $y$ and obtain new monomials
$x^2\zt ^2$ and $x\zt ^4$ instead. In the fibrations of the
geometries below it may happen that a specific fibration prefers 
a different parametrization then the one given above. However
the monomials always will have the identical scaling properties
as the ones given above.

\newsec{Six-dimensional heterotic $N=1$ vacua}
Let us proceed with six-dimensional dual pairs, that is F-theory
on Calabi--Yau three-fold $W_3$ versus the heterotic string on 
K3. This case has been studied from other various points of view in
\ref\AM{P.S. Aspinwall, \nup 496 (1997) 149; P.S. Aspinwall and D.R. Morrison,
\nup 503 (1997) 533.}%
\ref\Aii{P.S. Aspinwall, J. High Energy Phys. {$\us 4$} (1998) 19.}%
\ref\AD{P.S. Aspinwall and R.Y. Donagi, {\it
The Heterotic string, the tangent bundle, and derived
                  categories}, hep-th/9806094.}
\foot{In particular the result of local mirror symmetry is technically 
closely related to the stable degenerations of Calabi--Yau manifolds 
introduced in \FMW\ and discussed further in \AM.}.

After a brief discussion of the toric geometry for the elliptic
fibration in sect. 5.1. we turn to identifying the components of the
geometric moduli space in sect. 5.2. 
To demonstrate the method we will describe the construction of 
smooth bundles on K3 obtained from mirror symmetry in some detail in 
sect. 5.3. In sect. 5.4. we discuss gauge backgrounds that 
lead to non-perturbative gauge symmetry enhancements. In sect. 5.5.
we describe the geometric configuration for the tangent bundle. In 
sect. 5.6. we will derive new non-perturbative equivalences in 
six dimensional heterotic string compactification. The duality 
involves compactification on quite different K3 manifolds. In sect.
5.7. we discuss pairs of six dimensional compactifications on
mirror Calabi--Yau three-folds that become equivalent after 
further compactifying to three dimensions. In sect. 5.8 we  show
how the Higgs branches related to Coulomb branches in the dual
theory can be realized in terms of non-toric and non-polynomial 
deformations. In particular the physical spectrum is {\it not}
determined by the theoretical topological data of the Calabi--Yau
manifold, but by the number of deformations of a specific toric
realization of it.

\subsec{Toric geometry of the fibration}
As described in sect. 3.3. we have to fiber the local geometry 
$\cx W_2$ over a $\IP^1$. For the cases where $H\subset G_0$, with 
$G_0$ the heterotic gauge group, we can describe $\cx W_2$ as a local
patch of the K3 $W_2$ given by a polyhedron $\Ds_3$ as in Table 2.
In this fibration, $\Ds_3$ becomes the hypersurface $\cx H:\nus_{i,1}=0$
in a four-dimensional polyhedron $\Ds_4$ corresponding to a
Calabi--yau three-fold $W_3$. It remains to specify the
vertices of $\Ds_4$ which do not lie in $\cx H$.

There is no freedom in the choice of the base manifold; it is
just a $\IP^1$. In the toric polyhedron the vertices associated
to the $\IP^1$ are given by projecting along the fiber directions
\ref\KS{M. Kreuzer and H. Skarke, J. Geom. Phys. $\us 26$ (1998) 272.},
which maps a vector $\nus_i$ to its first entry. The vertices for
the toric variety $\IP^1$ are $\{(-1),(1)\}$. Thus we add vertices
with first entry $\pm1$.

Consider now the base of the {\it elliptic} fibration $\pi_F:W_3\to B_2$.
In the simplest case the base 
is a $\IP^1$ bundle over a base $\IP^1$ (with the base $\IP^1$
being the base of the elliptic K3 $W_2$), that is a Hirzebruch 
surface $\IFO_n$. The vertices of $\IFO_n$ are $\nus_i\in
\{(-1,0),(0,-1),(0,1),(1,n)\}$,
with the two relations between the $\nus_i$
corresponding to the two classes of $\IP^1$'s. The vertices with $\nus_{i,1}=0$
are already contained in the hyperplane $\cx H$. We add therefore two vertices
and obtain a polyhedron 
\eqn\tfpoly{
\Delta^*_4={\rm convex\  hull}\  
\{(0,{\nus_i}^\prime),\; (1,n,2,3),\; (-1,0,2,3)\}\; ,
}
where $\{{\nus_i}^\prime\}$ are the vertices of $\Delta^*_3$.

Furthermore, we can add two types of vertices corresponding to
non-perturbative dynamics of the heterotic string: $a)$ we can blow up the base
$\IFO_n$ of the elliptic fibration $\pi_F$ where the new moduli associated to
the blown up spheres correspond to non-perturbative tensor multiplets from 
five-branes in six dimensions \MVi,
$b)$ we can introduce
singularities in the elliptic fibration located at points on the base 
$\IP^1$. The new K\"ahler classes from reducible fibers 
correspond to non-perturbative gauge symmetries \MVi.

\subsec{The geometric moduli space}
Let us consider in more detail the precise meaning of the
map $f:\; W_{3}\to Z_2$ which gives
the heterotic manifold in terms of the polyhedron of the 
type IIA compactification.
The moduli space
of the type IIA compactification on $W_3$ has two sectors, 
the moduli space $\cx M_{HM}$ parametrized by the 
hypermultiplets and the vector multiplet moduli space $\cx M_{VM}$.
These spaces are in general decoupled due to the constraints
of $N=2$ supersymmetry up to subtleties explained e.g. in 
\ref\KMP{S. Katz, D.R. Morrison and M.R. Plesser, \nup 477 (1996) 105.}. 

The hypermultiplets of type IIA
on $W_3$ contain the string coupling, the complex structure of 
$W_3$ and the Ramond-Ramond (RR) fields. In the dual heterotic theory
on $Z_2\times T^2$ the hypermultiplets contain the geometric moduli
of $Z_2$ and the data of the bundle on $Z_2$. 
The complex structure of $W_3$ describes the 
geometry of $Z_2$ and part of the bundle data, namely the "spectral cover"
$C$ of $V$ or its generalizations defined by $\cx W_3$
for $H \neq SU(n)$. The RR moduli determine a line bundle $\cx L$
on $C$ \FMW. 

The 20 hypermultiplets that describe the geometry of K3
split into K\"ahler deformations and complex deformations
in a given algebraic realization of $Z_2$. In particular,
if $Z_2$ is elliptically fibered and has a global section, the
Picard lattice $Pic(Z_2)$ has at least rank two with a hyperbolic plane $U$
generated by the class of the section and the class of the elliptic 
fiber. We can therefore test only the part of the moduli space
of K3 compactifications with rank $Pic(Z_2)\geq2$ using the type 
IIA/F-theory picture. Actually we can argue that it is sufficient to
consider the case with rank $Pic(Z_2)=2$. Namely, since K\"ahler and 
complex structure deformations of a singularity are equivalent for K3,
we can always choose a complex deformation of a singularity to 
keep the rank of $Pic(Z_2)$ fixed\foot{The same will of course
not be true in the case of $n>2$, where K\"ahler and complex
structure deformations are not equivalent.}. Said differently, reaching
a singularity in the complex structure of $Z_2$, there is in general
no new branch in the moduli space corresponding to K\"ahler deformations
of this singularity. However there {\it can} be new branches in four
dimensions if
the singularity is associated with non-perturbative
gauge symmetries $\hx G$. In this case there is a new branch in the four
dimensional compactification on K3$\times T^2$  corresponding
to non-vanishing $\hx G$ Wilson lines on $T^2$ emanating from 
the locus of singular K3.

Keeping this in mind we can now make the meaning
of the map $f$ from the type IIA to the heterotic polyhedron  more precise.
The correspondence of the geometric hypermultiplet moduli space,
identifying the complex structure deformations of $W_{n+1}$
with the complex structure of $Z_n$ and the bundle data on it
has been discussed in the previous sections. It remains to 
assign the vector moduli, corresponding to K\"ahler moduli 
of the toric manifold $W_{n+1}$, to heterotic deformations.

In general, the K\"ahler deformations of the toric variety $W_{n+1}$
defined by a complex polyhedron $\Delta^*_{W_{n+1}}$ provide
always a complete resolution of the canonical singularities of  
birational equivalent models. In particular this implies also
the complete resolution of singularities for any divisor in $W_{n+1}$.
Let us assume that $\dzn$ appears not only as a projection but is 
a hyperplane in $\dwn$, that is $Z_n$ is a divisor in $W_{n+1}$. 
The K\"ahler resolution of $W_{n+1}$
provides a K\"ahler resolution of the divisor $Z_n$. However,
since the K\"ahler moduli of type IIA on $W_{3}$ correspond to vector 
multiplets whereas the K\"ahler deformations of the heterotic string on $Z_{2}$
are related to hypermultiplets, the K\"ahler blow up of the singularities
in the divisor $Z_2$ of $W_3$ is
{\it not} mapped to a K\"ahler blow up of $Z_2$. Rather
it corresponds to the above mentioned {\it four}-dimensional Coulomb branch of
non-perturbative gauge symmetries emanating from the singularity
in $Z_2$. So although the K\"ahler resolution of $W_3$
provides a K\"ahler resolution of a singularity of $Z_2$,
the correct heterotic picture is in terms of a Coulomb branch
of a  non-perturbative gauge symmetry compactified
on the {\it singular} K3 times $T^2$. Of course this interpretation is
only possible because of the equivalence of the moduli space
of K\"ahler deformations of  singularities in the elliptic fibration
of K3 manifolds with the moduli space of flat  bundles
on an elliptic curve! This is another, different consequence of the fact
that we used  elliptically fibered manifolds as 
the starting point for our geometric construction 
of flat  bundles. 

The fact that $\dzn$ appears in general only as a projection
rather than a hyperplane, fits nicely in the above picture. Since
$Z_2$ does not represent a divisor in $W_3$, the complete
K\"ahler resolution of $W_3$ does not necessarily provide a complete 
K\"ahler resolution of $Z_2$. This corresponds to the
situation with generic bundle $V$, 
where the singularity in $Z_2$ does not lead to 
a non-perturbative gauge symmetry and there is no new branch
in the moduli space.

\subsec{Smooth bundles}
We consider now in some detail configurations which correspond to smooth 
heterotic bundles at a generic point in moduli space. To specify the
theory we have to choose two bundles $V_1$, $V_2$ in the two $E_8$ factors and 
the integer $n$ which specifies the fibration of the K3 fiber $M_2$
over a further $\IP^1$. $n$ corresponds also to the way the total
instanton number $k=24$ is divided between the two $E_8$ factors: $k_1=12+n$,
$k_2=12-n$ \MVi. 

The geometry corresponding to the choice of $(V_1,V_2)$ and $n$ is as follows.
The bundle $(V_1,V_2)$ determines the K3 fiber $M_2$. As before
we take the structure group of $V_2$ to be trivial and concentrate
on the first $E_8$ factor. For a structure group $H_1$ we take
the corresponding K3 specified in Table 2. The instanton
number $k_1$ is encoded in the fibration of $M_2(H_1)$ over
the base $\IP^1$ with coordinates $(s,t)$ and corresponds to the choice 
of $n$ for the Hirzebruch surface $\IFO_n$~\MVi. We therefore consider
the polyhedron $\Ds_4$ of \tfpoly\ with $n=k_1-12$.

\subsubsec{$SU(N)$ bundles}
{}From the
polyhedron \tfpoly\ we obtain a Calabi--Yau three-fold $W_3(A_N)$.
The defining hypersurface is given by \batpol. Taking the local limit
of $p_{\Delta^*_4}$ and making our special choice of local coordinates 
we obtain the local three-fold geometry $\cx W_3$:
\eqn\Aan{\eqalign{
p_{\Delta_4^*}|_{\rm local}&=p_0+p_+\; ,\cr
p_0&=y^2+x^3+\zt^6f_{12}+y\zt^3h_6+x\zt^4h_8+x^2 \zt^2 h_4+y x \zt st\; ,\cr
p_+&=v\; (\zt^Nf_{k_1}+x\zt^{N-2}f_{k_1-4}+\dots+{x^{N/2}f_{k_1-2N}\brace
yx^{\ov{N-3}{2}}f_{k_1-2N}}).}}
Here $f_l$ is a generic polynomial of homogeneous degree $l$
in the variables $(s,t)$ while $h_l$ is of the restricted
form $h_l=s^l+\alpha_l  t^l$.
The interpretation of the three complex dimensional geometry $\cx W_3$ 
is very similar to the situation we encountered before: 
$v=0$ projects onto the K3 surface $Z_2:p_0=0$. This is the
K3 surface (dual 
to the manifold) on which the heterotic string is compactified.
Integrating
out the linear variable $v$ we obtain a one-dimensional 
geometry, the intersection $p_0=0\; \cap p_+=0$ which describes
a curve $C$ in $Z_2$. $C$ is the spectral curve which 
determines the $SU(N)$ bundle on $Z_2$ as in \FMW\BJPS.

The number of parameters of $p_+$ is $Nk_1-N^2+2$. Discarding
one parameter which can be absorbed in an overall rescaling
this agrees with the dimension of the moduli space of $A_N$
bundles of instanton number $k$ on K3 as determined by 
the index formula
\eqn\ifo{
{\rm dim}\; \cx M(H)=c_2(H)\; k-dim(H)\ ,}
which applies for simple $H$ and large enough $k$.
Strictly speaking the formula \Aan\ and similar formulae below for the
other groups are valid for the values of $N$ which appear in Table 2.
However it is worth noting that these formulae are valid for any large $N$
as well.
The only new aspect is that for large $N$ we have to consider non-compact
Calabi--Yau geometries as in \KMV.

\subsubsec{$SO(2N+1)$ bundles}
For the group $B_N$ (as well as for several other groups below)
we encounter the situation that the fibered geometry $W_3(H)$
includes monomials with identical scaling properties in 
addition to those encountered for $W_2(H)$.
In particular note that
the polynomial $p_+^1$ has degree $N$ in the variables 
$(y,x,\zt)\sim(\lambda^3 y,\lambda^2 x,\lambda \zt)$ but does not
contain all monomials of the appropriate weight. In the fibered
geometry $W_3(B_N)$, the monomials in $W_2(B_N)$ are multiplied
by general functions $f_l$ of the base variables $(s,t)$,
while the extra monomials are multiplied by restricted 
functions $h_l=s^l+\alpha_l t^l$. The general expression 
can be written as
\eqn\Bbn{\eqalign{
p^1_+=&\zt ^{N}f_{k_1}+\zt ^{N-3}yf_{k_1-6}+\eps yx^{\ov{N+1}{2}-2}f_{k_1-2N}
+(1-\eps) x^{\ov{N}{2}}f_{k_1-2N}+\cr
&\zt ^{N-2}xh_{k_1-4}+\zt ^{N-4}x^2h_{k_1-8}+
\zt ^{N-5}xyh_{k_1-10}+\zt ^{N-6}x^3h_{k_1-12}\cr
&+\dots +\eps \zt x^{\ov{N+1}{2}-1}h_{k_1-2N+2}
+(1-\eps)
\zt yx^{\ov{N}{2}-2}h_{k_1-2N+2}\; ,\cr
p^2_+=&\zt ^{2N-6}f_{2k_1-12}+\zt ^{2N-8}xf_{2k_1-16}+\dots+
x^{N-3}f_{2k_1-4N}\ ,}}
with $\eps=0\; (1)$ for $N$ even (odd).
The number of parameters is $(2N-1)k_1-(2N^2+N)+1$ as predicted
by the index formula. The $SO(7)$ case is again obtained from $SO(9)$
by dropping the second term $xf_{2k_1-16}$ in $p_+^2$.

\subsubsec{$Sp(N)$ bundles}
The $Sp(N)$ case can again be considered as a modding 
of the $SU(2N)$ bundle by the operation $y\to-y$. This
agrees with the result obtained from the geometric construction:
\eqn\Ccn{
p_+=v\; (\zt ^{2N}f_{k_1}+x\zt ^{2N-2}f_{k_1-4}+\dots+x^{N}f_{k_1-4N})\; .}
The number of parameters is  $(N+1)k_1-(2N^2+N)+1$ as predicted by
\ifo.

\subsubsec{$SO(2N)$ bundles}
For $H=SO(2N)$ we have again extra monomials because $p_+^1$
of $W_2(D_N)$ is not of the generic form. We find
\eqn\Ddn{\eqalign{
p^1_+=&\zt ^Nf_{k_1}+\zt ^{N-3}yf_{k_1-6}
+\zt ^{1-\eps}yx^{\ov{N+\eps}{2}-2}f_{k_1-2(N-1+\eps)}
+\zt ^\eps x^{\ov{N-\eps}{2}}f_{k_1-2(N-\eps)} +\cr
&\zt ^{N-2}xh_{k_1-4}+\zt ^{N-4}x^2h_{k1-8}+
\zt ^{N-5}xyh_{k_1-10}+\zt ^{N-6}x^3h_{k_1-12}+\dots+\cr
&\zt ^{3-\eps}yx^{\ov{N+\eps}{2}-3}h_{k_1-2(N+\eps-3)}+
\zt ^{2+\eps}x^{\ov{N-\eps}{2}-1}h_{k_1-2(N-2-\eps)}\; ,\cr
p^2_+=&\zt ^{2N-6}f_{2k_1-12}+\zt ^{2N-8}xf_{2k_1-16}+\dots+
\zt ^2x^{N-4}f_{2k_1-4(N-1)}\ .}}
with $\eps=0\; (1)$ for $N$ even (odd).
The number of parameters in $p_+^1$ and $p_+^2$ is
$(2N-2)k_1-(2N^2-N)+1$ in agreement with the index formula \ifo.

\subsubsec{$G_2$ bundles}
For $G_2$ bundles we obtain a geometry $W_3(G_2)$:
\eqn\Ccn{\eqalign{
p^1_+&= \zt ^3f_{k_1}+x\zt h_{k_1-4}+yf_{k_1-6}\; ,\cr
p^2_+&= f_{2k_1-12}\; .}}
The number of parameters is $4k_1-14+1$ in agreement with \ifo.

\subsubsec{$F_4$ bundles}
For $F_4$ bundles, the complex geometry $W_3(F_4)$ takes the
form
\eqn\Ffiv{\eqalign{
p^1_+&= \zt ^4f_{k_1}+x\zt ^2h_{k_1-4}+y\zt h_{k_1-6}+x^2f_{k_1-8}\; ,\cr
p^2_+&= \zt ^2f_{2k_1-12}+xf_{2k_1-16}\; ,\cr
p^3_+&=f_{3k_1-24}\; .}}
The number of parameters is $9k_1-52+1$. 

\subsubsec{$E_6$ bundles}
If we fiber the geometry $W_2(E_6)$ we obtain a three-dimensional 
manifold
\eqn\Eevi{\eqalign{
p_+^1=&\zt ^5f_{k_1}+\zt ^3xh_{k_1-4}+\zt ^2yh_{k_1-6}+\zt x^2f_{k_1-8}
+yxf_{k_1-10}\; , \cr
p_+^2=&\zt ^4f_{2k_1-12}+\zt ^2xf_{2k_1-16}+\zt yf_{2k_1-18}\; ,\cr
p_+^3=&\zt ^3f_{3k_1-24}\; .}}
The number of parameters is $12k_1-78+1$, as expected.

\subsubsec{$E_7$ bundles}
For $H=E_7$, the local geometry $W_3(E_7)$ takes the form
\eqn\Eevii{\eqalign{
p_+^1=&\zt ^5f_{k_1}+\zt ^3xh_{k_1-4}+\zt ^2yh_{k_1-6}+\zt x^2h_{k_1-8}
+yxf_{k_1-10}\; , \cr
p_+^2=&\zt ^4f_{2k_1-12}+\zt ^2xh_{2k_1-16}+\zt yf_{2k_1-18}
+x^2f_{2k_1-20}\; ,\cr
p_+^3=&\zt ^3f_{3k_1-24}+\zt xf_{3k_1-28}\; ,\cr
p_+^4=&\zt ^2f_{4k_1-36}\; .}}
The number of parameters is $18k_1-133+1$ and agrees with \ifo.

\subsubsec{$E_8$ bundles}
Finally we describe $E_8$ bundles over the K3 $Z_2:p_0=0$.
The geometry $W_3(E_8)$ reads
\eqn\Eeviii{\eqalign{
p_+^1=&\zt ^5f_{k_1}+\zt ^3xh_{k_1-4}+\zt ^2yh_{k_1-6}+\zt x^2h_{k_1-8}
+yxh_{k_1-10}\; , \cr
p_+^2=&\zt ^4f_{2k_1-12}+\zt ^2xh_{2k_1-16}+\zt yh_{2k_1-18}
+x^2f_{2k_1-20}\; ,\cr
p_+^3=&\zt ^3f_{3k_1-24}+\zt xh_{3k_1-28}+yf_{3k_1-30}\; ,\cr
p_+^4=&\zt ^2f_{4k_1-36}+xf_{4k_1-40}\; ,\cr
p_+^5=&\zt f_{5k_1-48}\; ,\cr
p_+^6=&f_{6k_1-60}\; .}}
The number of parameters is $30k_1-248+1$ as expected.

\subsec{Singular bundles and non-perturbatively enhanced gauge symmetries}
The heterotic vacua with smooth compactification manifold
$Z_H$ and generic $H$ bundle described so far have unbroken gauge
symmetries with group $G$, the commutant of $H$ in the perturbative
original ten-dimensional gauge group $G_0$. Let us investigate 
the conditions under which the heterotic string acquires extra 
massless degrees of freedom of non-perturbative origin. 
Note that the local mirror geometry $\cx W_3$, which defines the map of
moduli spaces, is part of an F-theory compactification on the
small fiber limit of $W_3$. We can therefore use the
F-theory knowledge of how to engineer non-perturbative gauge
symmetries to obtain a blow-up of $W_3$ corresponding to
a  non-perturbative gauge symmetry $G_{np}$ \MVi\Betal. This is
done by wrapping a seven-brane on the fiber $\IP^1$ of the base $B_F=\IFO_n$ of the
elliptic fibration $\pi^\prime_F:\ W \to  B_F$. The geometry 
$\cx W_3^\prime$ obtained from the blown up three-fold defines the
heterotic data in the same way as for the case of smooth bundles
and smooth $Z_2$. In the toric language this corresponds to adding a vertex
of type $b)$ in sect. 5.1. 
Let us first establish the following general result:
\vskip0.2cm

\leftskip .5cm \rightskip .5cm
\ni{\it  $(+)$
Consider the $E_8\times E_8$ string compactified on 
an elliptically fibered K3  with a singularity of type 
$\cx G$ at a point $s=0$ and a special gauge background $\hx V$. If
the restriction $\hx V_{|E_H}$ to the fiber $E_H$ at $s=0$ is
sufficiently trivial, the heterotic string acquires a 
non-perturbative gauge symmetry $G_{np} \supset \cx G$. }

\leftskip .0cm \rightskip .0cm
\ni 
Note that the above result is very similar to the case of
the type IIA string on singularities. In this case we 
know that type IIA on singularities of 2-cycles acquires
a non-perturbative gauge symmetry from D2-brane wrappings,
under the condition that  the background field $B$ vanishes
\ref\Asii{P. Aspinwall, \plt 357 (1995) 329.}.
In the heterotic string, not surprisingly, the vanishing condition includes 
also the gauge fields. It would be interesting to have a geometric 
interpretation of this gauge symmetry enhancement as in the brane
picture of the type IIA theory.

There are two comments in order. Firstly, since the geometric
data of $W_3$ describe only a subset of the heterotic 
moduli corresponding to the spectral cover and its generalization 
for other gauge groups, but not the extra information of a line bundle
$\cx L$ on it \FMW, the enhancement of gauge symmetry requires in addition
appropriate values for these non-geometric moduli.
Secondly, the non-perturbative gauge group is at least $\cx G$ for 
a gauge background of the  restricted type described below and
can be larger than $\cx G$, if additional restrictions 
on the behavior
of $\hx V$ in a neighbourhood of the singularity are imposed.
We will discuss such cases below.

The verification of the above claim is very simple using the
fact that the bundle is defined on the Calabi--Yau $Z_2:p_0=0$.
Addition of vertices of type $b)$ amounts to a singularity $\cx G$ of the 
Calabi--Yau manifold $W_3$ over a point on the base $\IP^1$. 
If $W_3$ is written in generalized Weierstrass form\foot{Note that we
use $(y,x,\zh)$ and $(y,x,\zt)$ to denote the homogeneous coordinates 
of the elliptic fiber of the $n+1$-dimensional Calabi--Yau $W_{n+1}$ 
and the $n$ dimensional Calabi--Yau $Z_n$, respectively.}:
\eqn\gwsf{
p=y^2+x^3+yx\zh a_1+x^2\zh ^2a_2+y\zh ^3a_3+x\zh ^4a_4+\zh ^6a_6\; ,}
the conditions for a singularity of type $\cx G$ have been
analyzed using the Tate's formalism in \Betal. The 
singularity at a point $s=0$ is determined by the powers
of vanishing of the coefficients $a_i\sim s^{n_i}$
specified by a vector $\bx n=(n_1,n_2,n_3,n_4,n_6)$.

Since the polynomial $p_0$ consists of a subset of the polynomials
in \gwsf, the coefficients $a_{i,0}$ of the generalized Weierstrass
form of $Z_2$ fulfil the same singularity condition as $p$,
so $Z_2$ has a $\cx G$ singularity.
Moreover from the explicit form of the bundle moduli space associated
to a structure group $H$ we see, that if the dependence of the 
$a_i$ is given by $\bx n$ then the leading behavior of the bundle is
shown\foot{We restrict to the subset of structure
groups $H$, for which the toric resolution results in a Weierstrass
form as in \gwsf.} in Table 3.
\vskip 0.3cm

\vbox{\eqn\bb{
\vbox{\offinterlineskip\tabskip=0pt\halign{\strut
\vrule\hfil~$#$~\hfil&\vrule#&
\hfil~$#$~\hfil&\hfil~$#$~\hfil&  \hfil~$#$~\hfil&
\hfil~$#$~\hfil&\hfil~$#$~\hfil&  \hfil~$#$~\hfil\vrule\cr
\noalign{\hrule}
H&&v&  v^2&   v^3&   v^4&   v^5&   v^6\cr
\noalign{\hrule}
SU(2)   &&\sx4 x&  -   &  -   &  -   &   -  &  -   \cr
SU(3)   &&\sx3 y&  -   & -    &  -   &  -   & -    \cr
SU(4)   &&\sx2 x^2&  -   &  -   &  -   &  -   &  -   \cr
SU(5)   &&\sx1 yx  &  -   &  -   &  -   &  -   &  -   \cr
SO(7)   &&\sx2 x^2 &\sx6 \zt ^2&  -   &  -   & -    &  -   \cr
SO(9)   &&\sx2 x^2 &\sx4 x  &  -   &  -   & -    &  -   \cr
SO(11)  &&\sx1 xy &\sx2  x^2&  -   &  -   & -    &  -   \cr
Sp(2)   &&\sx2 x^2 &  -   & -    &  -   & -    &  -   \cr
SO(8)   &&\sx2 x^2 &\sx4 x  &  -   &  -   & -    &  -   \cr
SO(10)  &&\sx1 yx &\sx4 x  &  -   &  -   & -    & -    \cr
G_2     &&\sx3 y&\sx6   &   -  &   -  & -    & -    \cr
F_4     &&\sx2 x^2&\sx4 x&\sx6   &   -  & -    & -    \cr
E_6     &&\sx1 yx &\sx3 \zt y&\sx6 \zt ^3&   -  &  -   &  -   \cr
E_7     &&\sx1 yx&\sx2 x^2&\sx4 x\zt &\sx6 \zt ^2& - &  -    \cr
E_8     &&\sx1 yx&\sx2 x^2&\sx3 y^3&\sx4x&\sx6 \zt &\sx6 \cr
\noalign{\hrule}
}}}
\leftskip .5cm \rightskip .5cm 
\noindent{\ninepoint  \baselineskip=8pt  
{{\bf Table 3:} Behavior of the gauge background  $\hx V$ 
near the K3 singularity.}}}

\leftskip .0cm \rightskip .0cm \ni 
For example consider the simplest case with a structure group
$H_1= SU(2)$. The Calabi-Yau threefold
$W_3(A_1)$ is defined by the polynomial 
\eqn\exaa{\eqalign{
&v^{-1}\zt ^6f_{24-k_1}+\cr
&v^0(y^2+x^3+yx\zt s t+x^2\zt ^2h_4+y\zt ^3h_6+x\zt ^4h_8+\zt ^6f_{12}) + \cr
&v\; (\zt ^6f_{k_1}+x\zt ^4f_{k_1-4}) \ ,
}}
where $h_i=s^i+t^i$ and $f_i$ is a generic degree $i$ polynomial as
before. To obtain a non-perturbative gauge symmetry $\hx G=SU(2)$ 
we blow up the locus $y=x=s=0$ in $W_3$:
\eqn\sht{
y=uy,\qquad x=ux,\qquad s=us\ .
}
This blow up is compatible with only a subset of the perturbations
in \exaa\ and we get a new manifold
\eqn\npgsii{
\eqalign{&v^{-1}s ^2\zt ^6f_{22-k_1}+\cr
&v^0(y^2+x^3u+x^2\zt ^2h_4+y\zt ^3s  
h_5+x\zt ^4s  h_7+yx\zt us t  + \zt ^6s ^2f_{10})+\cr
&v(\zt ^6s ^2f_{k_1-2}+x\zt ^4s  f_{k_1-5})}
}
Here $h_i$ and $f_i$ are as before apart from the fact that $s $
is replaced by $s  u$. Note that the heterotic manifold $Z_2:p_0=0$
has an $A_1$ singularity at $s =0$. The spectral cover has become
\eqn\spci{
p_+=s  (s  \zt^2f_{k_1-2}+xf_{k_1-5})=0
}
Let $S$ denote the class of the section of $Z_2$ and $F$ the class
of the generic elliptic 
fiber. The intersections are $S^2=-2$, $F\cdot S=1$, $F^2=0$.
The class of the spectral cover $\Sigma:p_+=0$
is then $2S+k_1F$ which is generically a smooth connected curve. 
The bundle $\hx V$ described by \spci\ instead corresponds to
a spectral cover with two components, $\Sigma=\Sigma_1+\Sigma_2$
with $[\Sigma_1]=F$ and $[\Sigma_2]=2S+(k_1-1)F$.

A physical interesting case which features a supposedly
self-dual heterotic string theory in six dimensions is 
the compactification of the heterotic $E_8 \times E_8$ string
with 12 instantons in each $E_8$ factor 
\ref\DMW{M.J. Duff, R. Minasian and E. Witten, \nup 465 (1996) 413.}
\ref\Bem{M. Berkooz et al., \nup 475 (1996) 115.}. 
It was argued in 
\Bem%
\ref\sensi{A. Sen, \nup 498 (1997) 135.}
that 
unhiggsing of a perturbative $SU(N)$ is dual to a non-perturbative
$SU(N)$ arising from $1+{N \over 2}$ $SO(32)$ small instantons
without vector structure at an $A_1$ singularity (using a
duality between $SO(32)$ and $E_8 \times E_8$ string on K3). Here
we see that the $E_8\times E_8$ picture of this gauge enhancement
is in terms of 
a compactification on a K3 with $A_{N-1}$ singularity, rather than $A_1$,
with a particular behavior of the vector bundle near the singularity
specified in Table 3\foot{For a related phenomenon in compactifications
on the tangent sheaf, see~\AD.}.

\subsec{Near the tangent bundle}
The class with structure group $H_1 =SU(2)$ described in 
eq.\exaa\ should contain also
the standard embedding with the gauge background identical to 
the spin connection. Since the $SU(2)$ bundle is embedded in
one $E_8$ factor we choose $k_1=24$. The spectral bundle
reads then
$$
C: \zt ^2f_{24}+xf_{20}=0 \; ,
$$
with the $46-1$ parameters
corresponding to the well-known 45 deformations of the tangent bundle 
$T$ of
K3. To get strictly the tangent bundle, first note that the 
restriction $V_{E_H}$ to the elliptic curve must be trivial 
due to the flatness of $T^2$. So we have to tune $f_{20}=0$.
Moreover the only pathologies of $T_{|E_H}$ appear at the singular
fibers of the elliptic fibration. These occur at the 24 zeros
of the discriminant $\Delta_{24}=4f_8^3+27g_{12}^2$ of $p_0$, where $f_8$ and
$g_{12}$ are related to the polynomials in \npgsii\ by a shift of
variables that puts $p_0$ into 
Weierstrass form, $y^2+x^3+x\zt ^4f_8+\zt ^6g_{12}$.
So we must choose $f_{24}={\rm const.}\ \Delta_{24}$. From the 
point of F-theory this geometry is identical to the one
with 24 small instantons sitting at the singular fibers of 
the fibration. This agrees with the result in \AD. There the authors
argue that the difference between the tangent bundle and
the small instanton configuration is in the non-geometric moduli
corresponding to RR fields in the type IIA theory. 

\subsec{Non-perturbative equivalences}

We now will study some applications
of the toric map $f:\; W_{3}\to Z_2 $
to investigate a certain class of six-dimensional heterotic theories 
with large non-perturbative groups and interesting non-perturbative
equivalences. 
To recap, we consider a type IIA/F-theory compactification
on an elliptically Calabi--Yau three-fold $W_3$ which also
has a K3 fibration whose fiber we denote by $W_2$. This is
dual to the heterotic string compactified on $Z_2$. In the toric 
polyhedron $\Ds_{W_3}$
$i)$ the fiber $W_2$ corresponds to a hyperplane $\cx H=\Ds_{W_2}$ in 
$\Ds_{W_3}$, $ii)$ the heterotic K3 $Z_2$ appears as the projection
$f:\; \Ds_{W_3} \to \Ds_{Z_2}$.

An interesting situation appears,
if the projection $f$ results in a hyperplane
$\cx H^\prime$, that is $W_3$ admits at the same time
a $Z_2$ fibration. The two K3 fibrations imply that we have two different
perturbatively defined heterotic theories in four 
dimensions, which are non-perturbatively equivalent\foot{The quite reverse
 situation is known to occur, in which two heterotic theories have the same
perturbative spectrum while non-perturbatively they are different~\ref\BKKM{P.
Berglund, S. Katz, A. Klemm, and P. Mayr, to appear.}.}. 

Moreover, if the two K3
manifolds defined by the hyperplanes 
$\cx H$ and $\cx H^\prime$ share the same elliptic fiber,
we can take the F-theory limit without interfering with the
equivalence\foot{To be precise, in order for the F-theory limit
to work, we have to require that not only $\cx H^\prime$ but 
also the hyperplane $\cx H:\; \nus_{i,1}=0$ coincides with a projection 
(in the first coordinate).}.
In this way we obtain two six-dimensional
heterotic compactifications on $Z_2$ and $W_2$ which are
non-perturbatively equivalent. These manifolds can be
at rather different moduli, one being highly singular
while the other being smooth, as we will see in the following example.

\subsubsec{24 small instantons on smooth K3}
Let us start with the simplest case corresponding to a heterotic theory
with 24 small $E_8$ instantons on a single point in a smooth K3 $Z_2$. The
perturbative gauge group is $E_8\times E_8$. The K3 fiber of the three-fold
$W_3$ is therefore the K3 $W_2$ with
$E_8\times E_8$ singularity described by the
polyhedron
\eqn\poli{
\Delta^*_{W^{E_8\times E_8}_2}=\chull{e_2,e_3,u_4,\tx u_4} \ ,}
where $W_2$ has an elliptic fibration with fiber 
$\Delta^*_{E_1}=\chull{e_2,e_3,f_1}$.
Let $\nu^{*(\al)}$ denote an $n+1$ dimensional vertex obtained from
an $n$ dimensional vertex $\nus$ by adding a zero at the $\al$-th position.
Since $W_2$ appears as a hyperplane $x_1=0$ in $\Delta^*_{W_3}$ in our
conventions, we obtain $\Delta_{W^{E_8\times E_8}_2}
^{*(1)}$ as the first piece of $\Delta^*_{W_3}$. We will refer to 
the K3 $W_2$ which is the fiber of the K3 fibration $W_3\to \IP^1$
as the "fiber K3".

The smooth K3 of the heterotic string appears as the projection in the
direction of the second coordinate. It is modeled by a polyhedron
\eqn\polii{
\Delta^*_{Z^{A_0}_2}=\chull{e_2,e_3,v_0,\tx v_0} \ .}
We add therefore $\Delta^{*(2)}_{Z_2}$ to $\Delta_{W_3}^*$.
We will refer to the "heterotic K3" $Z_2$, which is an elliptic
fibration over the base $\IP^1$ of the K3 fibration as the "base K3".

Finally we have to ensure convexity of $\Delta^*_{W_3}$. A discrete series
of solutions corresponding to the situation where the instantons
have been divided into two groups with $a+b$ and $24-a-b$ instantons
is provided by
\eqn\poliii{
\Delta_{W_3}^*=\chull{\Delta_{W^{E_8\times E_8}_2}
^{*(1)}\cup\Delta^{*(2)}_{Z^{A_0}_2}\cup
\tx\Delta^*_{A_0^{(1)}}\cup \tx\Delta^*_{A^{(2)}_0}}\ ,}
with 
\eqn\poliiib{\eqalign{
\tx\Delta^*_{A^{(1)}_1}&=\chull{(1,a,2,3),\; (1,-b,2,3)}\; ,\cr
\tx\Delta^*_{A^{(2)}_0}&=\chull{(-1,12-a,2,3),\; (-1,-12+b,2,3)}\ ,
}}
with $0\leq a,b \leq 12$. For simplicity we assume in the following
that both groups contain a non-zero number of small instantons. 

Note how simple the toric construction of the combined data
for the bundle and the manifold is. It is also easy to show that
the Calabi--Yau manifolds $W_3$ associated to $\Ds_{W_3}$ give indeed
the correct physics. Firstly, the hodge numbers are
$h^{1,1}=43(0)$, $h^{1,2}=43(22)$, where the number in
parentheses denotes the number $\delta h^{1,1}$ ($\delta h^{1,2}$)
of so-called non-toric (non-polynomial) deformations, which
are not available in the toric model. The $n_T+n_V=h^{1,1}-2$ 
vector and tensor multiplets are associated to the 16 vector multiplets
of $E_8\times E_8$, 24 tensor multiplets from
the 24 small $E_8$ instantons and the heterotic coupling (2 K\"ahler classes
corresponding to the volume of the elliptic fiber and the volume of
the base do not contribute to the vector and tensor multiplets \MVi).
The $n_H=h^{2,1}+1$ hypermultiplets arise from the 20 moduli from K3 and
2 moduli for the two positions of the two groups of fivebranes. 
The 22 missing complex structure moduli correspond naturally to the fact
that we have fixed 22 of 24 positions of the small instantons in the 
K3 $Z_2$.

There is a second elliptic fibration of  the K3 fiber $W_2$ due to the
hyperplane $\Delta^*_{E_2}=\chull{e_3,w_3,\tx w_3}$ corresponding to the
gauge group $SO(32)$\ref\caniii{P. Candelas and H. Skarke,
\plt 413 (1997) 63.}. 
Instead of $n_T=24$ extra tensors we have in this 
case an $Sp(a+b)\times Sp(24-a-b)$ 
gauge group from two groups of coincident $SO(32)$ five branes 
\ref\witsmi{E. Witten, \nup 460 (1996) 541.}
with matter in the $(\us{2k},\us{32})\oplus
(\us1,\us1)\oplus(\us{2k^2-k-1},\us{1})$ of each $Sp(k)\times SO(32)$
factor.

The non-perturbative gauge groups are determined by the intersections 
of the holomorphic 
two-cycles in $W_3$\foot{This is explained in detail in 
\KMV\ref\Lect{P. Mayr, {\it Geometric construction of N=2 gauge theories}, 
hep-th/9807096.}.}.
For the elliptic and K3 fibered manifolds used in the present context,
these intersections can be conveniently described by projections 
of the polyhedron $\Delta^*_{W_3}$ in the direction of the
elliptic fiber \ref\caniv{P. Candelas, E. Perevalov and G. Rajesh,
\nup 507 (1997) 445.}. 
For $W_3$ we find in this way indeed
a non-perturbative
gauge group $\hat{G}=\emptyset$ for the $E_1$ fibration and 
$Sp(a+b)\times Sp(24-a-b)$ for the $E_2$ fibration. 
Note that after having chosen
the perturbative gauge group corresponding to $W_2$ and the 
heterotic compactification $Z_2$, {\it the non-perturbative 
dynamics is completely
determined by convexity of the polyhedron $\Delta_{W_3}^*$},
with a discrete set of solutions corresponding to various
branches in the moduli space.

These are the first two interpretations of F-theory compactification on
$W_3$. Note that we have to shrink different elliptic fibers $E_1$ and
$E_2$ in the F-theory limit, so these theories are disconnected in
the small fiber limit in six dimensions and become equivalent only
in five dimensions by T-duality \ref\Gin{P. Ginsparg, \prv D35 (1987) 648.}.

\subsubsec{A non-perturbatively equivalent heterotic theory}
More interestingly, since the base K3 $\Delta^*_{Z^{A_0}_2}$,
corresponding to the heterotic compactification manifold,
does not only appear as a projection
but as a hyperplane, there is a second K3 fibration with fiber 
$Z_2$ which is itself elliptically fibered 
{\it with the same
elliptic fiber $E_1$} as the K3 fiber of the original K3 fibration. 
Therefore we obtain a theory in six dimensions
which is non-perturbatively equivalent to the heterotic string with 24 small 
instantons on on smooth K3. 

We interpret now the smooth K3 described by
$\Delta^*_{Z^{A_0}_2}$ as the fiber K3. Due to the absence of 
a singularity, the perturbative gauge group
must be trivial and therefore the bundle $V_0$ has
structure group $E_8\times E_8$ on the generic elliptic fiber.
On the other
hand, in the new K3 fibration, ${W^{E_8\times E_8}_2}$ has become the base K3.
The heterotic compactification manifold has therefore 
an $E_8\times E_8$ singularity. 
The perturbative $E_8 \times E_8$ gauge symmetry of the compactification
with small instantons is produced in the dual theory
purely by non-perturbative 
effects related to the singularities of the manifold and the bundle.
For $a=b=12$ we have 
therefore the following duality:

\noindent
\leftskip .5cm \rightskip .5cm 
{\it ($\dagger$) The $E_8\times E_8$ heterotic string compactified on a
smooth K3 with two groups
of 12 small instantons is non-perturbatively 
equivalent to compactifying on a 
K3 $p_0=0$ with $E_8\times E_8$ singularity with gauge bundle $V_0$. Here
\eqn\npdi{
p_0=y^2+x^3+yx\zt st +\zt ^6(s^7t ^5+s^6t ^6+s^5t ^7)\; .
}
The $E_8\times E_8$ bundle $V_0$ is specified by a geometry $\cx W$ 
as in eq. \Eeviii\  
of the special form
\eqn\npdib{\ninepoint
p_+=\tv(\zp5+yx)+\tv^2(\zp4+x^2)+\tv^3(\zp3+y)
+\tv^4(\zp2+x)+\tv^5\zp{}+\tv^6\; .\,}
with $\tilde{v}=vst$, $\zp{}=\zt st$ 
and a similar polynomial for $p_-$ for the other $E_8$ factor.
}

\noindent
\leftskip 0cm \rightskip 0cm 
\subsubsec{Small instantons on singular K3 manifolds}
The above situation can be very easily generalized to singular
K3 manifolds. Let us consider the case where we still start with
only small instantons, now on a singular K3 surface. This case
has been analyzed from various points of view in \ref\BI{J.D. Blum
and K. Intriligator, \nup 506 (1997) 199; \nup 506 (1997) 223.}\AM.
What is new is that with our understanding of bundles and manifolds
we will sometimes find non-perturbatively equivalent theories
involving a specific gauge background with non-trivial structure 
group on a different K3. 

The construction of the appropriate three-folds $W_3$ is 
by now standard using our kit of K3 polyhedra: to obtain
a compactification on a singular K3 we choose simply the
polyhedron of the base K3 to describe a K3 manifold
with a given singularity $G^\prime$. Requiring convexity of
the polyhedron $\Delta^*_{W_3}$ starting from $\Delta^{*(2)}_{Z^{G^\prime}
_2}\cup
\Delta^{*(1)}_{W^{E_8 \times E_8}_2}$ 
we obtain a set of discrete solutions corresponding
to a choice of positions for the small instantons. It is impressive
to observe that in fact all gauge groups derived in \BI\AM\ 
arise in the toric construction as a very simple consequence 
of the convexity of $\Delta^*_{W_3}$!

We will discuss only one further example, the
prototype case with an $A_1$ singularity in $Z_2$. This
will turn out to be interesting also from its surprising
relation to the compactification on the vector bundle 
near the tangent bundle. For $\Delta^*_{Z_2}$
we take the K3 with $A_1$ singularity
\eqn\polv{\Delta_{Z^{A_1}_2}^*=\chull{e_2,e_3,v_0,v_1,\tx v_0}.}
The solutions to the convexity with the 24 small instantons 
collected in two groups located at the singular point of $Z_2$ and
a smooth point, respectively, are given by
\eqn\polvi{
\Delta^*_{W_3}=\chull{\Delta^*_{W^{E_8\times E_8}_2}
\cup \Delta^*_{Z^{A_1}_2}\cup \tx \Delta^*_{A_1}\cup\tx \Delta^*_{A_0}}\ ,}
where the polyhedra $\tx \Delta^*_{A_1}$ and $\tx \Delta^*_{A_0}$ encode
the information about the small instantons at the singular
and smooth point of K3, respectively:
\eqn\polvii{\eqalign{
\tx\Delta^*_{A_1}&=\chull{(1,a,2,3),\; (1,a-2,1,2),\; (1,-b,2,3),\;
(1,-b+2,1,2)}\; ,\cr
\tx\Delta^*_{A_0}&=\chull{(-1,12-a,2,3),\; (-1,-12+b,2,3)}\ ,}}
with $2\leq a,b \leq 12$. This configuration 
corresponds to a collection of $A=a+b$ instantons on the $A_1$ singularity
and $B=24-A$ instantons on a smooth point in K3.

Determining the gauge symmetry from projecting along the 
fibers as in \caniv\ we find the gauge groups 
\eqn\shtx{\eqalign{
&E_8\times E_8 \times SU(2)_{np}^{A-3}\; ,\ n^\prime_T=24\; ,\cr
&SO(32)\times Sp(A)_{np}\times Sp(A-4)_{np}\times Sp(B)_{np}
\; ,\ n_T^\prime=1\; ,}}
for the elliptic fibers $E_1$ and $E_2$, respectively.
Moreover
$n_T^\prime$ denotes the number of tensor multiplets in addition
to the generic one.
This is in agreement with the results of \BI\ and \AM. For
the local geometry we have a chain of three ALE spaces
fibered over three ${\IP^1}$'s with intersections between the neighbors.
{}From the intersections we obtain  matter as in 
\ref\KV{S. Katz and C. Vafa, \nup 497 (1997) 146.} in addition to 
the gauge fields.

The heterotic manifold and the bundle $V$ on it can be obtained
in the now familiar way from \batpol\ and \lml. For the manifold
we obtain 
\eqn\stui{
p_0=y^2+x^3+x^2z^2h_4+yz^3sh_5+xz^4sh_7+z^6s^2f_{10}+yxz(st+t^2)\; ,}
which has an $A_1$ singularity at $s=0$. For the bundle we obtain 
\eqn\stuii{
p_+=vs^{A},\qquad p_-=v^{-1}t^B\; .}

Let us mention two cases which are physically interesting. The first
case is where all instantons are on the $A_1$ singularity, $A=24$. We will 
argue  in the next section 
that this theory is equivalent to a compactification on the 
deformation of the tangent bundle
of a smooth K3 after compactification on a $T^3$.
There is no non-perturbative dual in six dimensions since the projection
in the $\nu_{i,1}^*$ direction does not yield a K3 polyhedron.

As a second case consider $A=12$. This time there is a non-perturbative
dual and it is in terms of an $E_8\times E_7$ bundle on a K3 with 
$E_8\times E_8$ singularity with a similar structure as in eq. \npdib.
Quite generally it follows from the construction that the new 
duality implies: 
\vskip0.1cm
\leftskip 0.5cm \rightskip 0.5cm\ni
{\it $(\dagger \dagger)$ The heterotic string compactified  
on a K3 with $G^\prime$ singularity and with a certain gauge background
$V_0$ with structure group  $H$
is non-perturbatively equivalent to the heterotic string compactified
on a K3 with $G$ singularity and with a specific gauge background
$\tilde{V}_0$ with structure group  $H^\prime$}. 
\vskip0.1cm

\leftskip 0.cm \rightskip 0.cm\ni
Here $H$ $(H^\prime)$
is the commutant of $G$ $(G^\prime)$ in $E_8\times E_8$. 
Technically, this duality exists if the F-theory manifold is 
described by a polyhedron with  two K3 hyperplanes that coincide at the
same time with projections and intersect in a plane that corresponds to 
an elliptic curve. The bundles $V_0$ and $\tilde{V}_0$ are
determined by local mirror symmetry, as was done in the two examples above.

\subsec{F-theory on mirror manifolds}
It is instructive to look also at the heterotic theory corresponding
to type IIA/F-theory compactification on the mirror manifold $M_3$ of $W_3$. 
In particular, upon further compactification on a circle to three dimensions, 
the two type IIA theories on $M_3\times S^1$ and $W_3 \times \hx S^1$
should describe the same physics in virtue of  mirror symmetry of Calabi--Yau
three-fold and T-duality on the 
circle\foot{Here $\hx S^1$ denotes the T-dual of $S^1$.}
\ref\tdms{K. Intriligator and N. Seiberg, \plt 387 (1996) 513;\br
J. de Boer, K. Hori, H. Ooguri and Y. Oz, \nup 493 (1997) 101;\br
M. Porrati and A. Zaffaroni, \nup 490 (1997) 107;\br
K. Hori, H. Ooguri and C. Vafa, \nup 504 (1997) 147.}.
Assuming that $M_3$ is also elliptically fibered, we have a pair
of six-dimensional theories from F-theory compactification on 
$M_3$ and $W_3$.
We do not expect these two theories to be dual in any sense in 
six dimensions. However, 
these theories are equivalent after
compactification on a three torus to three dimensions. Using the toric 
map $f:W_3\to Z_2$ and some basic properties
of mirror symmetry of K3 we can easily derive the 
six-dimensional theories which become equivalent after further 
compactification.

Recall the claim $(*)$ concerning mirror symmetry of elliptically fibered 
K3 manifolds in sect. 3.2.
We can combine this result with the previous construction of Calabi--Yau
three-folds to obtain a map between six-dimensional F-theories. 
We use $X_2^H,Z_2^{\hx H}$ to denote the two K3 manifolds used 
as the K3 fiber and the model for heterotic K3, respectively.
Here $H$ and $\hat{H}$ denote the singularity type in the
elliptic fibration. 
Consider the 
family of Calabi--Yau manifolds $X_3(X_2^H,Z_2^{\hx H})$ which 
is constructed from choosing a convex closure of the vertices 
\eqn\sht{
\Delta^{*(1)}_{X_2^H}\cup\Delta^{*(2)}_{Z_2^{\hx H}}\subset \Ds_4\; ,}
as before. This manifold describes a heterotic theory on K3 $Z_2$
with $\hx H$ singularity with perturbative gauge group $H$. 
Depending on which vertices are added to those in \sht\ to
render $\Ds_4$ convex,
there can be also a non-perturbatively dual theory with the roles of $H$
and $\hx H$ exchanged. From the above we have that the mirror
manifold of $X_3(X_2^H,Z_2^{\hx H})$ is of type 
$\tx X_3(X_2^G,Z_2^{\hx G})$. Thus mirror symmetry relates the following
heterotic theories:

\vskip 0.5cm
{\baselineskip=12pt \sl
\goodbreak\midinsert
\centerline{\epsfxsize 4.5truein\epsfbox{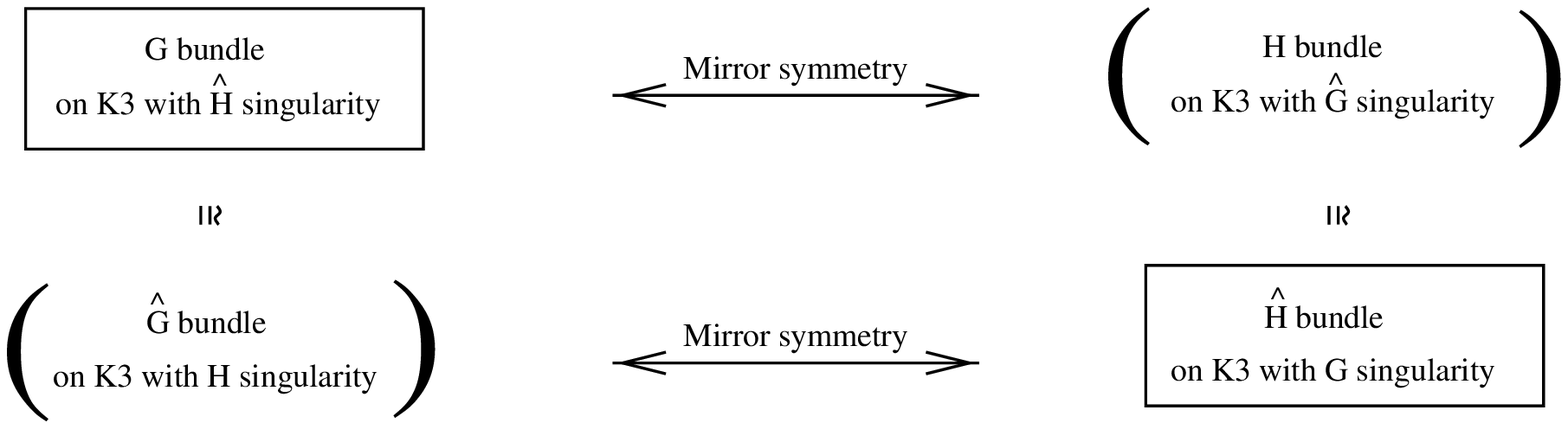}}
\leftskip 1pc\rightskip 1pc \vskip0.3cm
\noindent{\ninepoint  \baselineskip=8pt 
}\endinsert}\ni

Note that the formal mirror of the theory with $G$ bundle 
on K3 with $\hx H$ singularity
is the one with an $H$ bundle on K3 with $\hx G$ singularity
singularity (the two theories in the upper half of the diagram above).
However if we do not insist on the special case of double K3 fibrations,
the situation is different. The K3 fiber $X_2$ appears as a hyperplane
in the three-fold polyhedron, while the heterotic K3 $Z_2$ as a projection.
Moreover existence of a hyperplane (projection) in $\Delta^*$ 
corresponds to a projection (hyperplane) in $\Delta$.
In particular this means that the K3 fiber and heterotic manifold are
exchanged after the mirror transformation. 
The only generic K3 fibration in the mirror $\tilde{X_3}$ is the
one with a fiber that is the mirror of the {\it base} K3 of $X_3$.
The two relevant theories are denoted by a box in the above diagram.

As an example consider the case of a smooth $G$ bundle on a smooth K3,
that is $\hx H=\emptyset$. The compactification on the 
mirror is a heterotic theory with  24 small $E_8$ instantons on a 
$G$ singularity\foot{A relation between the instanton configurations
of the heterotic string theories dual to 
F-theory compactifications on a Calabi--Yau
3-fold and its mirror has been 
conjectured in \ref\PR{E. Perevalov and G. Rajesh, \prl 79 (1997) 2931.}
based on a comparison of topological data.}.

As a second example consider the compactification with 24 small
instantons on the $A_1$ singularity described in the previous section.
Using mirror symmetry we can show that the compactification on 
the mirror manifold describes 
an $SU(2)$ bundle with instanton number 24 on a smooth K3. In fact
from \batpol, the defining equations for the heterotic compactification
is $p=p_0+p_+ +v^{-1}$, with
\eqn\sht{\eqalign{
p_0&=y^2+x^3+\zt^6f_{12}+y\zt^3h_6+x\zt^4h_8+x^2 \zt^2 h_4+y x \zt st\; ,\cr
p_+&=v\; (z^2f_{24}+xf_{20})\; ,}}
in agreement with eq. \Aan. Note that 
from \shtx\ we have $h^{1,1}(W_3)=h^{1,2}(M_3)=64$
from the $16+21$ vector multiplets, $24$ tensors and three classes 
for the elliptic fiber and the two complex-dimensional base.
In the mirror these deformations correspond to $65=h^{1,2}+1$ 
hypermultiplet moduli, splitting in the 20 geometric hypermultiplets from 
K3 and $45=2\cdot24-3$ moduli for the $SU(2)$ bundle.
Similarly we have $h^{1,2}(W_3)=h^{1,1}(M_3)=18$, with the 19 
hypermultiplets of $W_3$ corresponding to the K3 moduli minus
one for the $A_1$ singularity. In the mirror, $h^{1,1}=18$ 
corresponds to the $15$
vectors from $E_8\times E_7$ and the generic three classes.

For the case $A=12$, we find that the compactification on the mirror
corresponds to a generic $k=12$ $SU(2)$ bundle on a smooth K3 plus
12 small instantons:
\eqn\sht{
p_+=v(z^2f_{12}+xf_8),\qquad p_-=v^{-1}f_{12}.}
Note that the instanton number $k$ is determined by the number of 
small instantons on the $A_1$ singularity. However, there is
a subtlety in the toric realization which follows from a simple 
counting of vector and hypermultiplets. For the theory on $W_3$
we get $52=3+16+9+24$ vector multiplets and $(20-1)+1$ hypermultiplets
for a configuration with the 12 instantons on the same smooth point. 
This corresponds to a toric manifold $W_3$ with hodge numbers
$h^{1,1}=52(0),\; h^{1,2}=30(11)$. The mirror with a $k=12$ $SU(2)$
bundle for a configuration
with the 12 instantons on a single smooth point should have
$n_V+n_T=30=3+15+12$, $n_H=42=20+21+1$ corresponding to
$h^{1,1}=30(0),\; h^{1,2}=52(11)$. On the contrary the mirror of $W_3$
has $h^{1,1}=30(11),\; h^{1,2}=52(0)$, that is a deficit of 11 vector/tensor
multiplets. This is related to the fact that the three-dimensional 
mirror symmetry,
which relates the theories after compactification on $T^3$, exchanges
Higgs and Coulomb branches. We now turn to a more detailed analysis of
the relation between a particular physical model and the toric representation.

\subsec{Higgs branches in toric geometry and self-dual string dynamics}

With the above understanding, we can refine the map between six-dimensional
theories and determine the non-trivial mapping between various 
Higgs and Coulomb branches. Interestingly we will find that 
the hodge numbers of the Calabi--Yau manifold do {\it not} in general
give the physical spectrum. We will still find complete agreement
between toric manifolds and physics using a subtle realization of 
Higgs branches in toric geometry in terms of 
non-toric and non-polynomial deformations $\delta h^{1,1}$ and
$\delta h^{1,2}$, respectively. Recall that these terms denote
the number of K\"ahler and complex structure deformations, which are
moduli of the general Calabi--Yau geometry $X$, but frozen in a
given toric realization. We find that the deformations
of the toric representation, $h^{i,1}_{eff}=h^{i,1}-\delta h^{i,1}$ 
agree precisely with the physical spectrum. 
In this sense the toric representation has more
physical meaning than the theoretical geometric data. This is very similar
to the F-theory duals
of the CHL string
\ref\CHL{S. Chaudhuri, G. Hockney and J.D. Lykken, \prl {75} (1995) 2264.}
and confirms 
the picture that the choice of toric representation with 
non-toric and non-polynomial deformations has an important physical
interpretation rather than being a technical 
subtlety~\ref\BKMT{P. Berglund, A. Klemm, P. Mayr and S. Theisen, 
{\it On type IIB vacua with varying coupling constant}, hep-th/9805189.}. 

Let us return to the example of $A$ small instantons on the $A_1$
singularity. In general, from \shtx, we would expect 
\eqn\shtii{
h^{1,1}=40+A\; ,\qquad h^{1,2}=18+1\; ,}
where the $+1$ denotes the extra position for the instantons on the
smooth point if $A\neq 24$, which we assume for the following. Rather than this
simple result, the polyhedra in eq. \polvii\ correspond to Calabi--Yau
manifolds $W_3$ with hodge numbers $h^{1,1}=40+A,\; h^{1,2}=18+B$.
To be consistent with \shtii\ there must be $\delta h^{1,2}=B-1$ non-polynomial
deformations of $W_3$. The actual story is much richer: there is
a chain of Calabi--Yau manifolds with this Hodge number but different
numbers of $\delta h^{1,1},\delta h^{1,2}$ and gauge symmetries 
shown in Tables 4,5.
The first member of Table 4 is the one with the correct hodge numbers
to describe the theory with $A$ instantons on the $A_1$ singularity and
$B$ instantons on a single smooth point, with the $\delta h^{1,2}=B-1$ 
positions corresponding to the $B-1$ fixed positions. 
Table 5 shows the same data for the compactifications
on the mirrors of those in Table 4. From the map
described in the previous section, the mirror of the first
theory in Table 4 
with $A$ instantons on $A_1$ singularity and $B$ instantons on 
a smooth point is a theory with an $SU(2)$ bundle of instanton
number $A$ and $B$ small instantons on a smooth point. This theory is
described by the last row in Table 5.

\vbox{
$$
\vbox{\offinterlineskip\tabskip=0pt\halign{
\strut
\vrule#&~$#$~\hfil\vrule&
\hfil~$#$~\hfil&~$#$~\hfil\vrule&
\hfil~$#$~\hfil&\hfil~$#$~\hfil\vrule\cr
\noalign{\hrule}
&\hfil G=G_{np}\times SO_{32}\; ,\ \ n_T^\prime =1\hfil &n_T^\prime
&\hfil G=G_{np}\times  E_8\times E_8 &h^{1,1}_{eff}&h^{1,2}_{eff}\cr
\noalign{\hrule}
 & Sp_A\times Sp_{A-4} \times Sp_B & 24 & (SU_2)^{A-3}  & 40+A & 19 \cr
 & \sut1\times Sp_A\times Sp_{A-4} \times Sp_{B-1} & 23 & \sut1 \times (SU_2)^{A-3}  & 40+A & 20 \cr
 & \sut2\times Sp_A\times Sp_{A-4} \times Sp_{B-2} & 22 & \sut2  \times (SU_2)^{A-3}  & 39+A & 21 \cr
 &\vdots&\vdots&\vdots&\vdots&\vdots\cr
 & \sut{B-1}\times Sp_A\times Sp_{A-4} \times Sp_{1} & A+1 & \sut{B-1}  \times (SU_2)^{A-3}  & 18+2A & B+18 \cr
 & \sut{B}\times Sp_A\times Sp_{A-4} & A & \sut{B}  \times (SU_2)^{A-3}  & 17+2A & B+18 \cr
\noalign{\hrule}
}}
$$
\leftskip0.5cm\rightskip0.5cm
\noindent{\ninepoint  \baselineskip=8pt 
{\bf Table. 4:}
Higgs branches emanating from type IIA compactification on $W_3$ with 
$A$ instantons on an $A_1$ singularity and $B=24-A$ small instantons on 
a smooth point. A superscript $SU_2^{(k)}$ denotes a level $k$ gauge group.}
}
\leftskip 0cm \rightskip 0cm \ni

\vbox{
$$
\vbox{\offinterlineskip\tabskip=0pt\halign{
\strut
\vrule#&~$#$~\hfil\vrule&
\hfil~$#$~\hfil&~$#$~\hfil\vrule&
\hfil~$#$~\hfil&\hfil~$#$~\hfil\vrule\cr
\noalign{\hrule}
&\hfil G=G_{np}\times SO_{28}\times SU_2,\; n_T^\prime =0\hfil &n_T^\prime
&\hfil G=G_{np}\times  E_8\times E_7 &h^{1,1}_{eff}&h^{1,2}_{eff}\cr
\noalign{\hrule}
 & \sut{B} & 0 & \sut{B} & 19 & 40+A \cr
 & Sp_1\times \sut{B-1}& 1& \sut{B-1} & 20 & 40+A\cr
 & Sp_2\times \sut{B-2}& 2& \sut{B-2} & 21 & 39+A\cr
 &\vdots&\vdots&\vdots&\vdots&\vdots\cr
 & Sp_{B-1}\times \sut{1}& B-1& \sut{1} & B+18 & 18+2A\cr
 & Sp_B& B&  & B+18 &17+2A\cr
\noalign{\hrule}
}}
$$
{\bf Table. 5:}
Higgs branches for the heterotic compactification on the mirror 
manifolds of those in Table 4.
}

Let us finally comment on the other theories with a different 
number of effective moduli. The polyhedra for the theories 
in Table 4 are obtained by successively dropping the edge vertex
of the line $\mu_i,\ i=1\dots B+1$ associated
to the $B$ small instantons on a single smooth point.
In the first step we loose one vertex from that line and catch a
new point $p=(0,-1,0,0)$ in the K3 polyhedron. In the $E_8 \times E_8$
language we have lost a tensor multiplet. The new point $p$
can not correspond to a tensor multiplet from a small instanton since
the associated divisor does not give a location on the base. 
Anomaly cancellation in
six dimensions as well as the fact that the size of the K\"ahler class
is blown down by the small fiber limit imply that the new branch 
exists in one lower dimension upon compactification on a circle.
Blowing down the divisor corresponding to the point $p$ we get a gauge
symmetry  enhancement to a 
new $SU(2)$ factor that is denoted by a tilde in the Tables 4,5. 
In the following, continuing dropping vertices from the line $\mu_i$,
at each step we loose one rank for the gauge group while the
self-intersection of the sphere associated to $p$ jumps by
$-2$. We interpret this effect as a higher level rank gauge
factor $\tilde{SU}(2)^{(k)}$.

In fact it is likely that these Higgs branches are associated to
extra matter, charged under the circle $U(1)$, which arises
in the $E_8\times E_8$ picture from self-dual string windings on
$S^1$. In particular, as we will argue in a moment, the $E_8$ five-brane
in six dimensions, when compactified on a torus, gives rise
to matter in the fundamental representation of the $SU(2)$ enhanced
gauge symmetry of the heterotic $E_8$ string
at $T=U$. While the vector 
bosons in this case arise from fundamental string windings that
become massless at $T=U$, we claim that the non-critical
string compactified on the same torus gives rise to hypermultiplets,
charged under the $SU(2)$, at the special radius. Whereas
the  massless vector states follow easily from the perturbative
world-sheet formulation of the heterotic fundamental string, a similar
phenomenon is somehow
surprising for the non-critical string; in particular there
is no known framework to describe the dynamics of the non-critical
string and the generation of massless states at $T=U$.

To see that the five-brane wrapped on the torus produces a fundamental
of the torus $SU(2)$, recall that geometrically, the five-brane
is associated to a blow up of the base, which is a Hirzebruch surface 
$\IFO_n$. $\IFO_n$ is a $\IP^1$ fibration over $\IP^1$, which in the
framework of geometric engineering of type IIA is well-known to give rise to 
a pure $SU(2)$ gauge theory 
\KKLMV. To add matter in this theory, one
blows up the $\IP^1$ fibration at a point of the base \KV. But
precisely the same is done in F-theory in six dimensions 
to get a tensor. Upon compactification on the torus,
which unifies the two pictures, the size of the fiber $\IP^1$
gets identified with the difference $T-U$ on the heterotic
side \ref\KLM{A. Klemm, W. Lerche and P. Mayr, \plt (1995) 313.}.
Thus the blow up corresponding to the $E_8$ fivebrane 
has become equivalent to the addition of a single doublet in the 
$SU(2)$ at $T=U$ of the heterotic string.

\newsec{$N=1$ supersymmetric vacua in four dimensions}

We proceed now with the most interesting case of $N=1$ supersymmetric
vacua in four dimensions. In sect. 6.1 we give a systematic
description of the bundle in terms of two line bundles $\cx L,\ \cx N$
on the base $B_{n-1}$. This determines essentially the $n+1$-fold
geometries $\cx W_{n+1}$, whose precise form can be obtained from 
the local mirror limit.  In sect. 6.2 we explain the relation of
these two bundles to similar bundles used in the construction
of Friedman Morgan and Witten. In sect. 6.3
we give explicit discussion of the case of toric four-folds\foot{Toric 
constructions 
of Calabi--Yau four-folds have been discussed in
\ref\KS{S. Kachru and 
E. Silverstein, \nup 482 (1996) 92.}%
\ref\FF{P. Mayr, \nup 494 (1997) 489.}%
\ref\KLRY{A. Klemm, B. Lian, S.S. Roan and S.T. Yau, \nup 518 (1998) 515.}%
\ref\Mohri{K. Mohri, {\it 
F theory vacua in four-dimensions and toric threefolds}, hep-th/9701147.}%
\ref\KS{M. Kreuzer and H. Skarke, J. Geom. Phys. $\us {26}$ (1998) 272.}
.} that
are fibered over the toric bases $\IP^2$ and $\bx F_n$, or blow ups thereof.
In sect 6.4 we describe the geometry
for compactifying on (deformations of) the tangent bundle
for any dimension $n$. In sect. 6.5 we describe how to construct
bundles on singular Calabi--Yau three-folds and describe new non-perturbative
dualities that are similar to the ones discussed in six dimensions.

\subsec{Description of the bundle}
Our construction of the local geometry $\cx W$ does not depend on the
dimension of the base $B_{n-1}$. As before we get the equation for the
mirror geometry from \batpol\ and taking the local mirror limit \lml, or
directly from \lmsr. Below we describe the outcome in a more economic
way in terms of the eight-dimensional geometries of sect. 4. This
general discussion will cover much of the information of the 
detailed result.

The two-fold
geometries of sect. 4 are defined
by the vanishing of the polynomial
\eqn\sht{p_{\cx W}=p_0+p_+=0\; .} 
The polynomial $p$ depends on 
four coordinates $(y,x,\zt,v)$. In the higher-dimensional fibration,
the coefficients of the polynomials in $p$ will become functions 
of the variables of the base $B_{n-1}$. More precisely, $(y,x,\zt,v)$
become sections of line bundles on $B_{m-1}$. These line bundles correspond
to the scaling relations $l^{(r)}$ of the polyhedron $\Ds$, as discussed in 
sect. 3.1. If we know
these scalings, we can determine the generic dependence of 
the function $f$ multiplying a term $y^a x^b \zt^c v^d$ of the
eight-dimensional local geometry. 

Naively we have to determine four different line bundles for the four variables
$(y,x,\zt,v)$. However we have one scaling relation \sri\ amongst them.
Let us call the line bundle associated to it $\cx O$. Moreover we have
assumed that the elliptic fibration has a section which gives a second
constraint.
In total we have therefore to specify  $2=4-2$ line bundles on $B_{n-1}$
to up-grade the two-dimensional local geometry $W_2$ to a $n+1$-dimensional
local geometry $\cx W_{n+1}$. 

Let us denote these two bundles by $\cx M$ and $\cx L$.
By definition, $y$ is a section of $\cx O^3$, $y\in \Gamma(\cx O^3)$.
Since we have the terms $y^2$ and $x^3$ appearing in $p$, we must have
$y\in \Gamma(\cx A^3\cx O^3)$, $x\in \Gamma(\cx A^2\cx O^2)$ and 
$p\in \Gamma(\cx A^6\cx O^6)$, for a line bundle $\cx A$. From the fact 
that 
\eqn\sht{
p_0=y^2+x^3+x\zt^4f+\zt^6g+\dots}
describes a Calabi--Yau manifold $Z_n$, with $f,g$ polynomials in the base variables,
we need to have 
$\cx A=\cx M \cx L$, with $\zt\in \Gamma(\cx M \cx O)$ and $\cx L$,
the anti-canonical bundle of $B_2$.
Moreover from $p_+\in \Gamma(\cx M^6 \cx L^6 \cx O^6)$ we have $v\in \Gamma
(\cx M^{5-N} \cx O^{6-N})$, with $N$ being the highest $z$ power in $p_+^1$.
This fixes almost completely the four-dimensional 
local geometry $\cx W_{n+1}$ in terms of $\cx W_2$:
\eqn\wfromw{\eqalign{
\cx W_2 &\to \cx W_{n+1}\cr
y^a x^b \zt^c v^d a_{c,d} &\to y^a x^b \zt^c v^d f_{c,d}\; ,\cr
f_{c,d} &\in \Gamma(\cx M^d\cx L^{c})\; .
}}
Note that the fact that $a=0,1$ and $3a+2b+c=6$ implies that the
coefficient function $f_{c,d}$ has only two indices, which is equivalent
to our previous statement that the definition involves  
only two independent line bundles.
Eq. \wfromw\ gives a general definition of the line bundle $\cx M$
for any fibration, also for non-toric bases. We will give explicit
expressions for toric bases in sect. 6.3.

\subsec{More on the line bundle $\cx M$ and stability of $V$}
Let us compare the above result from mirror symmetry with the general 
structure of the moduli space of holomorphic stable vector bundles 
found in \FMW. Friedman, Morgan and Witten described 
holomorphic stable bundles on elliptically fibered Calabi--Yau
manifold $Z$ by fibering the data $(E,W^r)$ holomorphically
over a complex base $B$. Here $W^r$ is a short-hand notation for the
weighted projective space $\IW^r_{s_0,\dots,s_r}$ predicted by Loojienga.
The projective spaces $W^r$ fit into a 
holomorphic bundle $\us W^r$ over $B$. If $s:B\to \us W^r$ is
a section, the homogeneous coordinates $\tilde{a}_i$ of $\us W^r$
pull back to sections
\eqn\omb{
\tilde{a}_i\in H^0(B,\cx N^{s_i}\otimes \cx L^{-d_j}),}
where $\cx L$ is the anti-canonical bundle of $B$, 
$\cx N$ a line bundle on $B$,
the $d_j$ are the degrees of the independent Casimir operators of 
the group $H$ and $s_i$ the Dynkin indices as above. 
Part of the data of $V$ are defined by choosing a section of $\us W^r$.
\def\coom{c_1(\cx N)}
The line bundle $\cx N$ is an important characteristic of $V$ and its
first Chern class $\eta=\coom$ is
closely related to the higher Chern classes of $V$\FMW\ref\BA{B. 
Andreas, {\it On vector bundles and chiral matter 
in N=1 heterotic compactifications}, hep-th/9802202;
G. Curio, \plt 435 (1998) 39.}. 

Our local mirror construction has given a similar answer for
the structure of the bundle: the relation between the geometry
and the associated topological data is essentially 
determined by the two line bundles $\cx L$ and $\cx M$. Of
course the two descriptions should agree and in fact it is 
easy to see that eqs. \wfromw\ and \omb\ imply
\eqn\fmwvsbm{
\cx N = \cx M \cx L^6.}

Eq. \wfromw\ gives a general definition
of the line bundle $\cx N$ in terms of the toric construction for all structure
groups $H$ as well as singular configurations such as sheafs and
geometric singularities. In particular we see that $\cx N$ is 
described in toric terms as a simple linear relation $l^{(\cx N)}$ between the
vertices of the toric polyhedron $\Ds$. As we will see in a moment, 
convexity of a polyhedron $\Ds$ that describes a fixed structure group
imposes a constraint on the possible values of $\eta=c_1(\cx N)$.

\subsec{Toric bases}
To get a four-dimensional theory by considering the F-theory
compactification associated to
the local mirror limit of the type IIA geometry we fiber the
local geometries $\cx W_2$ over a two complex dimensional base $B_2$.
$B_2$ will also be the base of the elliptic fibration $\pi_H:\; Z_3\to B_2$ 
of the Calabi--Yau three-fold $Z_3$ on which the bundle $V$ is defined.
A toric representation can be given for the cases $B_2=\IP^2$ or $\IFO_n$,
or a series of blow ups thereof. With these two choices for $B_2$ 
the toric polyhedra $\Ds_{B_2}$ are the convex hull
of the vertices
\eqn\piiv{
\vbox{\offinterlineskip\tabskip=0pt\halign{\strut
\hfil~$#$~\hfil&\vrule#&
\hfil~$#$~\hfil&\vrule#& 
\hfil~$#$~\hfil\cr
\IP^2&&\mus_i\in \Ds_{B_2}&&K\cr
\noalign{\hrule}
s&&(1,0)&&1\cr t&&(0,1)&&1\cr u&&(-1,-1)&&1\cr
}}}
\eqn\fnv{
\vbox{\offinterlineskip\tabskip=0pt\halign{\strut
\hfil~$#$~\hfil&\vrule#&
\hfil~$#$~\hfil&\vrule#& 
\hfil~$#$~\hfil&\hfil~$#$~\hfil\cr
\IFO_n&&\mus_i\in \Ds_{B_2}&&K_1&K_2\cr
\noalign{\hrule}
t&&(0,1)&&1&-n\cr s&&(0,-1)&&1&\cr 
t^\prime&&(1,0)&&&1\cr s^\prime&&(-1,n)&&&1\cr
}}}
Here we have also indicated the scaling relations between the
vertices and associated coordinates on them. These coordinates
transform as sections of the line bundles defined by the
scaling relations\foot{A letter $K$ denotes the class 
of the line bundle $\cx K$.}, e.g. $t^\prime \in \Gamma(\cx K_1^{0}\otimes \cx K_2^1)$.

As before, the three-fold 
base $B_3$ of the 
elliptic fibration $\pi_F:\; W_4\to B_3$ will determine
the topological properties of the gauge bundle $V$ on the
Calabi--Yau $Z_3$. The base $B_3$ is a $\IP^1$ bundle over $B_2$
with the following toric data. From $B_2=\IP^2$ we get
a $\IP^1$ bundle that we call $\tIF_k$:
\eqn\ftk{
\vbox{\offinterlineskip\tabskip=0pt\halign{\strut
\hfil~$#$~\hfil&\vrule#&
\hfil~$#$~\hfil&\vrule#& 
\hfil~$#$~\hfil&\hfil~$#$~\hfil\cr
\tilde{\IFO}_k&&\mus_i\in \Ds_{B_3}&&K_0&K\cr
\noalign{\hrule}
z&&(0,0,1)&&1&-k\cr
w&&(0,0,-1)&&1&\cr
s&&(0,1,0)&&&1\cr 
t&&(1,0,0)&&&1\cr
u&&(-1,-1,k)&&&1\cr
}}}
The integer $k$ specifies the $\IP^1$ bundle over $\IP^2$.
Similarly we get from $B_2=\IFO_n$ a $\IP^1$ bundle $\tIF_{k,m,n}$:
\eqn\ftk{
\vbox{\offinterlineskip\tabskip=0pt\halign{\strut
\hfil~$#$~\hfil&\vrule#&
\hfil~$#$~\hfil&\vrule#& 
\hfil~$#$~\hfil&\hfil~$#$~\hfil&\hfil~$#$~\hfil\cr
\tilde{\IFO}_{k,m,n}&&\mus_i\in \Ds_{B_3}&&K_0&K_1&K_2\cr
\noalign{\hrule}
z&&(0,0,1)&&1&-k&-m\cr
w&&(0,0,-1)&&1&&\cr
t&&(0,1,0)&&&1&-n\cr 
s&&(0,-1,k)&&&1&\cr
t^\prime&&(1,0,0)&&&&1\cr
s^\prime&&(-1,n,m)&&&&1\cr
}}}
with the two integers $k,m$ specifying the $\IP^1$ fibration.

The four-fold $W_4$ will be defined by a polyhedron $\Ds_5$ which 
is obtained by joining the vertices of the base $B_2$ and the
vertices of the K3 polyhedron $\Ds_H$ \ktl\ which describes the $H$ bundle:
\eqn\pff{\Ds_5=\chull{\Delta^{*(1,2)}_H\cup\tilde{\Delta}^\star_{B_3}\;}\; .}
Here the superscript $(1,2)$ denotes adding zeros at the first
and second position of the K3 vertices in \ktl\ 
and $\tilde{\Delta}^\star_{B_3}=\{\nus: \nus=
(\mus,2,3)\; ,  \mus \in \Ds_{B_3}\}$. 

We can add three kinds of vertices to the polyhedron (some of which 
may be enforced by the convexity condition);
$a^\prime)$ blow ups of the $\IP^1\to B_2$ fibration. These are
similar to case $a)$ in six dimensions and 
correspond to non-perturbative tensor multiplets from five-branes 
in the heterotic dual; $b^\prime)$ extra
singularities of the elliptic fibration located on the base $B_2$
related to non-perturbative gauge dynamics of the heterotic dual; 
$c^\prime)$ blow ups of the base $B_2$. Note that  due to the eight-dimensional
equivalence the base $B_2$ is visible
to both the heterotic string and the F-theory. Thus 
a modification of $B_2$ is common to both theories.

As for the two line bundles appearing in \wfromw, we have 
\eqn\cbb{\eqalign{
\cx L(\IP^2)&=3K=3D_s\; ,\cr
\cx L(\IFO_n)&=2K_1+(2-n)K_2=2D_s+(2-n)D_{s^\prime}\; ,}}
with $D_x:\; x=0$. The bundle $\cx M$ is determined by 
the transformation properties of, say,  $y$ under the rescalings of $\Ds$.
Consider first $\tIF_k$. Including homogeneous coordinates $(y,x,\zh)$
for the elliptic fiber, the polyhedron $\Ds_5$ has scaling relations
\eqn\ftkf{
\vbox{\offinterlineskip\tabskip=0pt\halign{\strut
\hfil~$#$~\hfil&\vrule#&
\hfil~$#$~\hfil&\hfil~$#$~\hfil&\hfil~$#$~\hfil&
\hfil~$#$~\hfil&\hfil~$#$~\hfil&
\hfil~$#$~\hfil&\hfil~$#$~\hfil&\hfil~$#$~\hfil\cr
&&y &x &\zh&z &w &s &t &u\cr
\noalign{\hrule}
\cx O&&3 &2 &1  &  &  &  &  &  \cr
\cx K&&3\la&2\la &   &-k & &1  &1  &1  \cr
}}}
with $\la=3-k$. From the scaling properties of $y$ and $\cx L=\cx K^3$
it follows that $\cx M(\tIF_k)=-k\; \cx K$. The relation between the 
global toric coordinates in \ftkf\ and the coordinates $(y,x,z,v)$
of the local geometry in this case can be seen to be
\eqn\lvgc{
(y,x,\zt,v)_{local}=(y,x,\zh z w, w z^{-1} (\zh z w)^{6-N})_{global}\; ,}
with the result that $v\in\Gamma(\cx M^{5-N}\cx O^{6-N})$ as promised.
Similarly we find $\cx M(\tIF_{k,m,n})=-k\; K_1-m\; K_2$. So
for the bundle $\cx N$ in \fmwvsbm\ we have
\eqn\etaclasses{\eqalign{
\tIF_k:\ &\cx N=(18-k)\cx K\ ,\cr
\tIF_{k,m,n}:\ &\cx N=(12-k) \cx K_1+(12-6n-m)\cx K_2\ .}}

\subsec{Heterotic $(2,2)$ compactification in $10-2n$ dimensions}
Using the F-theory/heterotic map presented above it is straightforward
to provide the F-theory manifold $X_{n+1}$ which corresponds to 
a heterotic compactification on the tangent bundle in $10-2n$ dimensions.
The heterotic gauge bundle takes values in the structure group $SU(n)$.
A generic deformation of the tangent bundle is obtained by fibering the
relevant K3 manifold in Table 2. The tangent bundle itself is a
highly degenerate version of the general $SU(n)$ bundle for the
same reasons as in sect. 5.5: the tangent bundle on the generic
elliptic curve is trivial, so the bundle (or rather sheaf) $V$ of
the heterotic compactification must be trivial away from the
discriminant locus of the fibration $\pi_H:Z_n \to B_{n-1}$.
The only pathologies of $V$ appear at the discriminant locus of
the fibration. In this way we get the following "theorem":

\leftskip0.5cm\rightskip0.5cm\noindent{\it
Let the heterotic string be compactified on the elliptically fibered 
Calabi--Yau $Z_n$, $\pi_h:Z_n\to B_{n-1}$ described in Weierstrass form by 
\eqn\tbi{
p_H=y^2+x^3+x z^4f+z^6g=0\; ,}
where $f$ and $g$ are sections of $\cx L^4$ and $\cx L^{6}$,
respectively.
The spectral cover of the two components of the 
sheaf $V$ in the gauge group $E_8\times E_8$
is described by the equations
\eqn\tbii{\eqalign{
p_+=z^n\Delta\; , \qquad p_-=1\; ,}}
where $\Delta$ is the discriminant of the elliptic fibration $\pi_H$,
$\Delta=4f^3+27g^2$. This is dual to F-theory compactified on the 
Calabi--Yau $n+1$ fold $X_{n+1}$ given by
\eqn\tbiii{
p_F=y^2+x^3+x(\zh z w)^4f+(\zh z w)^6g+\zh^6z^5w^7\Delta+\zh^6z^7w^5\; .}}

\leftskip0cm\rightskip0cm
\noindent
Here $(z,w)$ denote the variables parametrizing the base $\IP^1$
of the elliptically fibered K3 fiber of the K3 fibration $\pi_F:X_{n+1}\to
B_{n-1}$.
Eqs. \tbii\ and \tbiii\ fix the K3 fibration structure
of $X_{n+1}$ completely. In particular, from \tbii\ and
\wfromw\ we see that $f^{0,0,6,-1}\in \Gamma(\cx M^{-1}\cx L^6)$
must be a section of the trivial bundle so $\cx M = \cx L^6$.
E.g., the standard embedding for the two base geometries $\IP^2$ and $\IFO_n$ 
discussed in the previous section arises from F-theory on an
elliptic fibration over $\tIF_{18}$ and $\tIF_{12,6(2-n),n}$,
respectively.

\subsec{Bundles on singular manifolds and 
non-perturbative dualities in four dimensions}
As in six dimensions, there are non-trivial 
non-perturbative dualities in four dimensions implied by
the local mirror construction (and corresponding dualities
will also exist in lower dimensions). Let $\Ds_5$ describe the
polyhedron associated to a Calabi--Yau manifold $W_4$. 
There is a codimension two hyperplane $\cx H=\Ds_3$ 
associated to an elliptically fibered  K3, $W_2$. The elliptic fiber
is described by a hyperplane $\Ds_2(E)$ in $\Ds_3$. 
The $W_2$ fibration of $W_4$ describes $H$ bundles on a 
Calabi--Yau three-fold
$Z_3$ given by a projection in the direction of the section
of $W_2$.  In our conventions the 
hyperplane is given by $x_1=x_2=0$ and the projection is in the
third direction. A non-perturbative dual in four dimensions exists if
$i)$ there is a second codimension two hyperplane $\cx H^\prime$ that
describes an elliptically fibered K3 manifold $W_2^\prime$ with a section and
the same elliptic fiber $E$ as $W_2$: $\cx H \cap \cx H^\prime=\Ds_2(E)$; $ii)$
the projection in the direction of the section of $W^\prime_2$
results in a reflexive polyhedron for a Calabi--Yau three-fold $Z^\prime_3$.
We have:
\vskip0.1cm
\leftskip 0.5cm \rightskip 0.5cm\ni
{\it $(\dagger \dagger \dagger)$ Let the heterotic string be compactified  
on a Calabi--Yau three-fold with $G^\prime$ singularity and 
with a certain gauge background with structure group  $H$
such that the toric data $\Ds_5$ fulfil conditions $i)$ and $ii)$.
Then there exists a 
non-perturbatively equivalent compactification
on a Calabi--Yau manifold 
with $G$ singularity and with a specific gauge background
with structure group  $H^\prime$}. 
\vskip0.1cm

\leftskip 0.cm \rightskip 0.cm\ni
Again $H$ $(H^\prime)$
is the commutant of $G$ $(G^\prime)$ in $E_8\times E_8$. 

Note that as in six dimensions it is very easy to describe
$H$ bundles on Calabi--Yau manifolds with $G^\prime$ singularity
from our construction. If $\Ds_3$ is the polyhedron 
associated to $H$ bundles as in sect. 3.2 and $\Ds_4$ 
the polyhedron associated to the Calabi--Yau manifold $Z_3$
with a $G$ singularity, we consider polyhedra 
$\Ds_5$ that provide a convex closure of the vertices   
\eqn\sht{\eqalign{
&(0,0,\mu_i),\  \mu_i \in \Ds_3,\cr
&(\rho_{i,1},\rho_{i,2},a_i,\rho_{i,4},\rho_{i,5}),\ \rho_i \in \Ds_4}}
with $a_i$ integers that determine the topological data of the bundle.
As a simple example consider a theory with a smooth $E_8\times E_8$
bundle on a Calabi--Yau manifold $\pi:Z_3\to \IFO_0$ with an $A_1\times A_1$
singularity in the elliptic fibration. The polyhedron is given by
\eqn\sht{
\Ds_5=\Delta_{A_0}^{*(1,2)}\cup \Delta_{Z_3}^{*(3)}}
where we use the same notation as in sect. 5. Explicitly, we have 
\eqn\sht{\eqalign{\ninepoint
\Delta_{A_0}^{*(1,2)}&=\chull{\ninepoint(0,0,\pm 1,2,3),(0,0,0,-1,0),
(0,0,0,0,-1)},\cr
\Delta_{Z_3}^{*(3)}&={\rm convex \ hull}
{\ninepoint(\pm 1,0,0,2,3),(0,\pm1,0,2,3),(0,\pm1,0,1,2)\brace
(0,0,0,-1,0),(0,0,0,0,-1)}}}

These data correspond  to a toric Calabi--Yau manifold with 
$h^{1,1}(\delta h^{1,1})=6(0)$, 
$h^{1,2}=0$, 
$h^{1,3}(\delta h^{1,3})=2130(0)$ and $\chi=12864=0\; {\rm mod}\; 24$. 
Note that we have three non-perturbatively equivalent theories
corresponding to a K3 fiber given by $x_a=x_b=0$ and a projection to 
a Calabi--Yau manifold $Z_3$ in the $c$ direction:
\vskip0.2cm
\item{-}$(a,b,c)=(1,2,3)$: An $H=E_8\times E_8$ bundle on an elliptically
fibered Calabi--Yau $\pi:Z_3\to \IFO_0$ with $A_1\times A_1$ singularity;
\item{-}$(a,b,c)=(1,3,2)$: An $H=E_7\times E_7$ bundle on an elliptically
fibered Calabi--Yau $\pi:Z_3\to \IFO_0$ without singularity;
\item{-}$(a,b,c)=(2,3,1)$: An $H=E_8\times E_8$ bundle on an elliptically
fibered Calabi--Yau $\pi:Z_3\to \IFO_0$ with $A_1\times A_1$ singularity.
\vskip0.2cm
\ni
One also easily can  construct dualities where all three theories
are on different Calabi--Yau manifolds, and moreover such that the 
latter are fibered over bases $\IFO_n$ with different $n$.

\newsec{Phenomenological $N=1$ $d=4$ F-theory/heterotic vacua}
Let us finally apply our framework to construct dual pairs 
of F-theory/heterotic vacua with phenomenologically interesting 
gauge bundles. In particular, we will consider $SU(N)$ bundles
in one $E_8$ factor 
for $1\leq N\leq6$ on elliptically fibered Calabi--Yau three-folds with  
bases $B_2=\IP^2$ or $\IFO_n$ corresponding to unbroken gauge groups 
$E_8,\; E_7,\; E_6,\; SO(10),\; SU(5)$ and $SU(3)\times SU(2)$\foot{Note that
these gauge groups can be further broken by the non-geometric moduli.}.

\subsec{Moduli and spectra of the four-dimensional theories}
The moduli of F-theory on Calabi--Yau 4-folds 
and the heterotic string dual on Calabi--Yau 3-fold appear in several
ways in
the different formulations according to whether they arise
geometrically or non-geometrically,  perturbatively or non-perturbatively.

The perturbative moduli of the heterotic string are 
\item {$i)$} the complex structure of $Z_3$,
\item {$ii)$} the K\"ahler structure of $Z_3$,
\item {$iii)$} the moduli of the "spectral cover" $C$,
\item {$iv)$} the moduli of a line bundle $\cx L$ on the resolved fiber
product $C \times_{B_2} Z_3$.

Clearly, $i)$ and $iii)$ correspond to the complex structure of $W_4$.
There is one further complex structure modulus of $W_4$ that
we lost in the local mirror limit. It is related to the K\"ahler
class of the heterotic elliptic fiber. 

The moduli $iv)$ should be identified with instanton bundles on coinciding 
seven branes \BJPS. Note that an instanton inside a seven brane
corresponds to an Euclidean three brane wrapped on the interior
part of the seven brane
\ref\mdi{M.R. Douglas, {\it Gauge fields and D-branes}, hep-th
9604198.}. The specification of $\cx L$
includes the behavior on the exceptional divisors of the resolution
of the fiber product $C \times_{B_2} Z_3$. Moreover, continuous
moduli arise from non-trivial elements in $H^1(C)$. They
correspond to elements in $H^{1,2}(W_4)$ \FMW. In addition one
can twist the Jacobian $H^{1,2}(W_4,\bx R)/H^{1,2}(W_4,\bx Z)$ by elements
in $H^{2,2}(X_4,\bx Z)$ \FMW\CD. This twisting contributes to the 
gravitational anomaly 
\ref\SVW{S. Sethi, C. Vafa and E. Witten, \nup 480 (1996) 213.}%
, as has been discovered in the M-theory
context in terms of four-form flux 
\ref\BS{K. Becker and M. Becker, \nup 477 (1996) 155.}.
The equivalent contribution in F-theory arises from
the instanton bundle in the seven brane due to the world-sheet 
couplings \ref\GHM{M. Green, J.A. Harvey and G. Moore,
Class. Quant. Grav. $\us{14}$ (1997) 47}. 

\subsubsec{Non-perturbtaive gauge symmetries}
As for the remaining K\"ahler moduli of $Z_3$, the $h^{1,1}(B_2)$
moduli in the base maps to the same moduli in F-theory, since the
base is common to both theories in the compactification from eight
dimensions.
We are left with the K\"ahler moduli of $Z_3$ that arise neither
from the base $B_2$ nor from the ellipitc fiber, that is reducible
fibers from singularities of the elliptic fibration. 
We will now argue 
that they are equivalent to deformations of non-perturbative
gauge dynamics of the heterotic string.

Let us first consider a simple case, where we have a
non-perturbative gauge symmetry $G$ in the F-theory manifold $W_4$
from a $G$ singularity    in the elliptic fibration over a divisor $D$  
with codimension
one in the base $B_2$. The singularity can be
described in the Weierstrass form \gwsf\ in terms of the vanishing
of the coefficients $a_i$. As argued in sect. 5.4, 
if $D$ is toric, then the vanishing conditions on the $a_i$ are shared 
by the polynomial $p_0$ specifying the heterotic manifold; that
is, $Z_3$ has (at least) a $G$ singularity. 
The K\"ahler blow up of this singularity
in $Z_3$ can not correspond to a similar blow up in $W_4$, since
in the F-theory limit these moduli are frozen to zero. However, there
are other natural moduli associated to the $G$ singularity in $W_4$,
namely the moduli of a non-trivial instanton bundle on the
coinciding seven branes associated to the singularity over $D$.
They are the only possible candidates for the dual of the
heterotic moduli. Note that since the heterotic non-perturbative 
gauge symmetry $G$ is associated to the singular geometry, we indeed
expect that the blow up moduli are charged with respect to
$G$ and can be used to break the gauge symmetry.

In the reverse direction we can argue as follows. Assume we have
a smooth $E_8\times E_8$ bundle on an elliptic fibration $Z_3$
with a singularity $G$ over a curve $C$ in $B_2$. From 
the above identifications there are no perturbative moduli available in
the F-theory which could correspond to the K\"ahler blow up
of $G$ in $Z_3$. The only moduli that are not ruled out by 
the perturbative argument are moduli from instantons on 
coinciding seven branes on a "non-perturbative" divisor $D$.

There is another possible scenario starting from the singular
heterotic manifold in a case where there are no moduli
from instantons on coinciding seven branes associated to non-perturbative 
gauge symmetries: there is an obstruction to blow up 
in the heterotic theory. This is reminiscent of a similar situation
in the formulation of heterotic compactifications in terms
of linear sigma models, where it can happen that one cannot
extend a gauge background defined on the singular manifold
to another one on the resolved manifold within this 
framework~\ref\bt{P. Berglund and S. Theisen, unpublished.}.

\goodbreak
\subsubsec{Non-perturbative five-branes}
To complete our dictionary, there are two further types of moduli of 
non-perturbative origin in the heterotic theory. Firstly, there 
are $h^{1,1}_{5B}=h^{1,1}(B_3)-h^{1,1}(B_2)$ moduli of the F-theory manifold
that have no correspondence in the perturbative heterotic theory.
{}From the analogy to five-branes in six dimensions they give rise to 
non-perturbative tensor multiplets in four dimensions associated to non-critical
strings in the four-dimensional theory\foot{Such strings have been
suggested as a new mechanism for low energy supersymmetry breaking 
in \FF.}. A toric vertex $\nu$ associated to $h^{1,1}_{5B}$ is one of the 
type $a^\prime)$ specified in sect. 6.1 and the divisor (curve) in $B_2$ 
asspciated to $\nu$ is the supersymmetric 2-cycle on which the five brane is 
wrapped. This is clear in the case of $B_2=\IFO_n$ which can
be considered as a compactification of a six-dimensional theory
on the base $\IP^1$ of $\IFO_n$. The six-dimensional tensor
multiplet gives rise to an antisymmetric  tensor $B_{\mu\nu}$  
and a scalar $\phi$ in four dimensions which combine to an
$N=1$ linear multiplet. 

Secondly, if the instanton bundle 
does not balance the gravitational anomaly from the curvature
along $B_2$, there are additional five branes 
wrapped on the elliptic fiber $E_H$ of $Z_3$ \FMW. 
There is a $U(1)$ field and three complex scalars 
associated to the five brane, which specify a position on $B_2$
and a Wilson line on $E_H$. They correspond to the positions 
of a D3 brane on $B_3$ in the F-theory context\foot{In the M-theory
picture,
apart from the contribution from four-flux on $H^{2,2}(W_4)$ and 
membranes filling space-time, there
appears to be a further contribution to the gravitational anomaly
from five branes wrapped around 3-cycles in $W_4$ \FF.}.

\subsubsec{Non-perturbative gauge symmetries from non-toric bases}
Another non-perturbative effect arises from elements $h^{1,2}(B_3)$
which give rise to vector multiplets in four dimensions. This
can be argued from considering a compactification on $S^1$ and comparison 
with M-theory, which leads to the following contribution of 
the hodge numbers of $W_4$ to the four-dimensional 
fields \Mohri\foot{See also ref.
\ref\ACL{B. Andreas, G. Curio and D. L\"ust, \nup 507 (1997) 175.}.}:
\eqn\spectrum{\eqalign{
n_V&=(h^{1,1}(W_4)-h^{1,1}(B_3)-1)+h^{1,2}(B_3)\; ,\cr
n_C&=h^{1,1}(B_3)+(h^{1,2}(W_4)-h^{1,2}(B_3))+h^{1,3}(W_4)\; .}}

In our case, since $B_3$ is toric, $h^{1,2}(B_3)=0$ and there are
no contributions from $h^{1,2}(W_4)$ to vector multiplets.
In the toric framework that we use these hodge numbers can be
determined from the toric polyhedra $\Ds_5$ and $\Delta_5$ using
Batyrevs formula \ref\bathdg{V.V. Batyrev, J. Alg. Geom. $\us 3$ (1994) 493;
V.V. Batyrev, D.I. Dais, Topology $\us {35}$ (1996) 901;
V.V. Batyrev and L.A. Borisov, Invent. Math. $\us {126}$ 
(1996) 183.}:
\eqn\bathodge{\eqalign{
h^{1,i}=&\delta^{1,i}\Big(l(\Ds)-(n+2)-\sum_{\codim \theta^\star=1}
l^\prime(\theta^\star)\Big)+
\sum_{\codim\theta^\star=i+1}l^\prime(\theta^\star)l^\prime(\theta)
+\cr
&\delta^{n-1,i}\Big(l(\Delta)-(n+2)-
\sum_{\codim \theta=1}l^\prime(\theta)\Big)\ .}}
Here $\theta$ denotes a face of $\Delta$, $\theta^\star$ the dual face
and moreover $l$ and $l^\prime$ are the number of total and interior 
points of a face, respectively.

\subsec{Fibrations over $\IP^2$}
Consider the case where the heterotic manifold is given 
by a fibration over $\IP^2$.
The compactification is specified by the choice of $N$ for the $SU(N)$
gauge bundle together with the integer $k$ that determines
the base of the elliptic F-theory fibration $\tIF_k$. The
requirement that the singularities in the ellipitic fibration
are not worse than $E_8$ implies
\eqn\sht{|k|\leq 18\; .} Note that this constraint gives a bound 
on the class $\coom$ as can be seen from \etaclasses.
We will discover a similar bound for non-trivial structure groups in a moment.

As in six dimensions, the number $k$ determines how the
total Chern class $c_2(V)=c_2(V_1)+c_2(V_2)+[W]=c_2(TZ)$ of the 
gauge background is distributed among the $E_8$ factors. Here $[W]$
is the class of the fivebranes that are 
wrapped around the elliptic fiber \FMW\
and are needed to ensure tadpole cancelations
\SVW.

The starting point for type IIA manifolds $W_4$ is the polyhedron 
$\Ds_5$ described in eq.\pff. In this convention, $k=18$ corresponds
to a configuration near the  standard embedding with the non-trivial
gauge background embedded completely in one $E_8$ factor. 
Since the bundle in the second $E_8$ factor is trivial, we expect a
gauge group $G_N\times E_8$ with $G_N$ the commutant of $SU(N)$
in $E_8$ and the extra $E_8$ factor arising from the second $E_8$.
More generally, similar situations arise for large $k<18$ 
as a reflection of the fact that maximal breaking associated 
to the gauge background embedded in the second $E_8$ with fixed
Chern classes terminates at a non-trivial extra gauge group $G^{(2)}\subset E_8$
(ingoring extra breaking due to non-geometric Higgs moduli).
The gauge groups $G^{(2)}$ and the extra vertices $\mu_i$ which we add to $\Ds_5$
to describe the toric resolution of the associated singularities
are shown in Table 7.

\vskip .5cm
\vbox{
$$
\vbox{\offinterlineskip\tabskip=0pt\halign{
\strut
\vrule\hfil~${#}$~\hfil&\vrule\hfil~${#}$~\hfil&
\vrule\hfil~${#}$~\hfil\vrule\cr
\noalign{\hrule}
k&G^{(2)}&\mu_i\cr
\noalign{\hrule}
4&A_1&(0,0,1,2,3),\; (0,0,1,1,2)\cr
5,6&G_2&(0,0,2,2,3)\cr
7,8,9&F_4&(0,0,3,2,3)\cr
10,11,12&E_7&(0,0,4,2,3),\; (0,0,3,1,2),\; (0,0,2,0,1),\; (0,0,1,0,0)\cr
13,\dots,18&E_8&(0,0,6,2,3)\cr
\noalign{\hrule}}}
$$
\leftskip .5cm \rightskip .5cm 
\noindent{\ninepoint  \baselineskip=8pt  
{{\bf Table 7:} Singularities and toric resolution for $B^{(2)}=\IP^2$}}.}

\leftskip .0cm \rightskip .0cm \ni
Similarly if $k$ is small, the "instanton number" in the first 
$E_8$ bundle might not be sufficient to support the $SU(N)$ bundle
for large $N$. This gives a restriction on possible combinations 
of $N$ and $k$. In other words, there is a lower bound on $\eta=c_1(\cx N)$
for a given structure group $SU(N)$.

In Table B.1 we collect
the geometric data that enter the
spectrum of the heterotic compactifications\foot{Note
that there is an irrelevant sign switch in our notation of $k$ as compared
to sect. 6.3.}.
In particular we observe,
that for a structure group $H=SU(N)$, we need $k\geq 3N$ which means
in view of  \etaclasses\ that 
\eqn\sht{\eta = c_1(\cx N) \geq N c_1(\cx L)\ .}

The spectrum as determined by \spectrum\ is as follows: the
vector multiplets follow from the gauge group $G_N\times G^{(2)}$.
In many cases, for $N=1$ there is an extra $\tilde{SU}(2)$ factor of the 
same origin as in the discussion of Higgs branches in six dimensions
that explains the exceed of one in $h^{1,1}$. 
The hodge number
$h^{1,2}$, which is
expected to agree with the genus of the spectral cover, can be
quite large for the manifolds under consideration.
\hfil

\subsec{Fibrations over $F_n$ using deformation of the standard embedding}
Let us now consider the case with $Z_3$ an elliptic fibration over
$B_2=\IFO_n$. 
Requiring as before that the singularities of the ellipitc fibration 
are not worse than $E_8$ one obtains the restriction
\eqn\sht{
| k |\leq 12,\qquad -\big(12+n(6+k)\big)\; \leq\;  m\; \leq 12+n(6-k) \ ,
}
which again defines a lower bound on $c_1(\cx N)$ in \etaclasses.
We will consider only the case close to the
standard embedding with the gauge
background embedded in the $SU(N)$ bundle of a single $E_8$ factor,
corresponding to $k=12$ and $m=12-6n$.

The case $B_2=F_n$ is different from the $B_2=\IP^2$ models discussed
above in that for $|n|\geq 3$, the elliptic fibration over $B_2$ will 
have unavoidably singularities. Moreover the possible intersections of these
singularities with those describing the perturbative gauge symmetries
will lead to extra singularities above these intersections that need
to be resolved in the Calabi--Yau four-fold $W_4$. 
In general this requires a case by case study of
the actual singularities and their intersections. However we can 
give quite a canonical description for the following interesting combinations
of $n$ and $N$: $i)$ the trivial $SU(1)$ bundle for all $n$; 
$ii)$ $SU(N)$ bundles for $|n|\leq 6$. 

Let us first consider the singularities associated to large $|n|$,
in a heterotic compactification on the trivial $SU(1)$ bundle. 
The toric resolution of singularities in $W_4$ is described by
adding the following vertices $\tilde{\nu}^\star$ to $\Ds_5$ in \pff:\br
 
\vbox{
$$
\vbox{\offinterlineskip\tabskip=0pt\halign{
\strut
\vrule\hfil\ ${#}$\ \hfil&\vrule\hfil\qquad ${#}$ \qquad \hfil\vrule\cr
\noalign{\hrule}
n&\tilde{\nu}^\star\cr
\noalign{\hrule}
-12,\dots,-9&(0,-6,42,2,3)\cr
-8,-7&(0,-4,30,2,3)\cr
-6,-5&(0,-3,24,2,3)\cr
-4&(0,-2,18,2,3)\cr
-3&(0,-1,8,2,3)\cr
-2,\dots,2& - \cr
3&(0,1,-4,2,3)\cr
4&(0,2,-6,2,3)\cr
5,6&(0,3,-12,2,3)\cr
7,8&(0,4,-18,2,3)\cr
9,\dots,12&(0,6,-30,2,3)\cr
\noalign{\hrule}}}
$$
\leftskip .5cm \rightskip .5cm 
\noindent{\ninepoint  \baselineskip=8pt  
{{\bf Table 8:} Toric resolution for a trivial
bundle on the fibration $\pi:Z_3\to\IFO_n$}}.}

Taking the convex hull of these vertices leads to Calabi--Yau
four-folds with the topological properties\foot{The Calabi--Yau
manifolds associated to the polyhedra $\Ds_5$ as defined above 
have identical hodge numbers for $n$ and $-n$ and we can therefore
restrict to one sign for $n$.} collected in Table B.2.
For $k>2$ there are additional non-perturbative degrees of freedom
that contribute to $h^{1,1}$:
\eqn\sht{\eqalign{
n=-3&:\ \ntp=4,\ G_{np}=SU(2)\; ,\cr
n=-4&:\ \ntp=6,\ G_{np}=SU(2)^2\times G_2\; ,\cr
n=-5,-6&:\ \ntp=10,\ G_{np}=SU(2)^2\times G_2^2\times F_4\; ,\cr
}}
Note that these spectra are precisely what we expect from 
$k=4,6,8$ small $E_8$ instantons on $E_8$ singularity, respectively.

For $N>1$ a toric resolution of the intersections of singularities
can be described by further adding the vertices $\rho$ to $\Ds_5$:\br

\vbox{
$$
\vbox{\offinterlineskip\tabskip=0pt\halign{
\strut
\vrule\hfil\ ${#}$\ \hfil&\vrule\hfil\qquad ${#}$ \qquad \hfil\vrule\cr
\noalign{\hrule}
n&\rho\cr
\noalign{\hrule}
-6,-5&(0,-1,4,2,3)\cr
-4&(0,-1,6,2,3)\cr
-3,\dots,3& - \cr
4&(0,1,-6,2,3)\cr
5,6&(0,1,-8,2,3)\cr
\noalign{\hrule}}}
$$
\leftskip .5cm \rightskip .5cm 
\noindent{\ninepoint  \baselineskip=8pt  
{{\bf Table 9:} Toric resolution for 
$SU(N)$ bundles on the fibration $\pi:Z_3\to\IFO_n$.}}}

\ni
The topological data for these cases are collected in 
Table B.3.\foot{For
$|n|=4$ and $N=2$ we have omitted the vertex $\rho$, since addition
of the vertices $\tilde{\nu}^\star$ provides already a valid resolution.}

The non-perturbative degrees of freedom contributing to $h^{1,1}$ are
\eqn\sht{\eqalign{
n=7,8&:\ \ntp=14,\ G_{np}=\tilde{SU}(2)\times SU(2)^2\times G_2^3\times F_4\; ,\cr
n=9,\dots,12&:\ \ntp=22,\ G_{np}=\tilde{SU}(2)\times SU(2)^4\times G_2^4
\times F_4^2\times E_8\; ,\cr
}}
These spectra are almost identical to  what we expect from 
$k=9,10$ small $E_8$ instantons on an $E_8$ singularity, respectively.
We interpret the modifications as deformations of these configurations.

\newsec{Conclusions}
We have seen how mirror symmetry can be used to define vector bundles
on Calabi--Yau $n$-folds $Z_n$ and implies F-theory/heterotic duality
at the classical level.
In particular the construction allows for a very systematic 
identification of a dual pair realized in toric geometry,
consisting of a Calabi--Yau $n+1$-fold $W_{n+1}$ for F-theory compactification
and a Calabi--Yau $n$-fold $Z_n$ together with a family of vector
bundles on it defining a heterotic theory.
While the construction proves F-theory/heterotic duality at the
classical level, we would really like to go on to a comparison of 
quantum corrected properties, both to verify duality at the quantum
level and also to calculate quantities that are difficult to access
in one of the two theories, assuming duality. There are two 
obvious candidates for quantum corrected quantities accessible by
the geometric framework.
Firstly, we can use mirror symmetry to 
compute correlation functions of 
two-dimensional topological sigma models
\ref\wittft{E. Witten, Comm. Math. Phys. $\us{117}$ (1988) 353;
{\it Mirror manifolds and topological field theory}. In:
Essays on Mirror Manifolds (ed. S.-T. Yau),
Hong Kong: Int. Press, 1992, pp. 120-158.} and compare to
correlators in the space time theory. The general framework of
mirror symmetry of Calabi--Yau $n$-folds has been developped in
\ref\GMP{B.R. Greene, D.R. Morrison and M.R. Plesser, \cmp 173 (1995) 559.} 
and has been implemented into toric geometry in \FF, with
an identification of the relevant correlators in M-theory\foot{For
a relation to space time quantities in the two dimensional type IIA 
compactification, see \ref\WFF{W. Lerche, J. High Energy Phys. $\us{11}$ (1997) 4;
Nucl. Phys. Proc. Suppl. $\us {68}$(1998) 252.}.}.
In general the correlators as computed by the topological sigma model
are further modified by quantum corrections that do not arise
from Euclidean fundamental string world-sheet wrappings. However, favorable
situations in which these extra corrections are absent should exist and
could be used  for a comparison of heterotic and F-theory moduli
spaces at the quantum level\foot{An interesting comparison of 
quantum corrected correlators in eight dimensions appeared in
\ref\LS{W. Lerche and S. Stieberger, 
{\it Prepotential, mirror map and F theory on K3}, hep-th/ 9804176.}.}.

A second important quantity of
the four-dimensional $N=1$ theory is its superpotential for 
the moduli fields. In M-theory compactified on $W_4$ to three dimensions,
the superpotential arises from Euclidean five-brane wrappings on 
six-cycles $D$ with special topological properties
\ref\witsp{E. Witten, \nup 474 (1996) 343.}, with an
instanton action proportional to the volume $V(D)$ of $D$.  For smooth 
$D$ the condition is that its arithmetic genus $\chi(D)$ is equal to one.
This framework has been
used in
\ref\consp{S. Katz and C. Vafa, \nup 497 (1997) 204;\br
C. Vafa, Adv.Theor.Math.Phys. $\us 2$ (1998) 497.}
to calculate the superpotential of $N=1$
field theories and their compactifications to three dimensions.
In the toric Calabi--Yau manifolds used in the present
paper, the divisors $D$, $\chi(D)$ and $V(D)$ can be
systematically determined from intersection calculus \FF.
It would be very interesting to apply and extend these toric methods
in phenomenologically interesting $N=1$ $d=4$ theories
constructed in this paper.
\vskip 0.2cm

\noindent
{\bf Acknowledgements}:
P.M. thanks S. Katz and C. Vafa for many valuable discussions
related to \KMV.
The work of P.B. and P.M. was supported in part by
the Natural Science Foundation under Grant No. PHY94-07194. P.B. would
also like to acknowledge the Aspen Center for Physics and LBL,
Berkeley for hospitality during the course of this work.

\appendix{A}{The local mirror limit}\br\noindent
To justify the limit \lml\ we have to show that it corresponds
to the image in complex structure
of the local limit in K\"ahler moduli space under the action
of mirror symmetry. To decouple the K\"ahler deformations of the
local resolution of an $H$ singularity in the elliptic fibration 
over a point $z=0$ on the base $\IP^1$, 
we require first an infinite volume for the base
and then concentrate on the flat local neighborhood of $z=0$.

The K\"ahler moduli $t_r,\; r=1,\dots,h^{1,1}$ 
measure the volumes of holomorphic curves
in $M_{n+1}$ with the volume form given by the K\"ahler form. In toric
geometry, a non-trivial homology class corresponds to a linear 
relation $l^{(r)}$ between the vertices $\nu_i$ of the polyhedron 
$\Delta_{M_{n+1}}$:
\eqn\applmli{
\sum_{i} l^{(r)}_i\nu_i=0\; .}
In the absence of non-toric deformations there are 
$h^{1,1}$ relations of this type.
The specific choice of a basis of such linear relations 
which generates the K\"ahler cone is provided by the so-called Mori vectors.
  
Let us consider $n=1$, the case $n>1$ being completely analogous. 
For the K3 manifolds described in sect 3.2, the vertices $\nu_i$
can be divided into three groups $N_\Sigma,\ \Sigma\in\{0,+,-\}$ 
according to the sign of the first entry. The vertices with a zero
entry form the polyhedron $\Delta_E$ of the elliptic fiber, while
the vertices with positive (negative) first entry correspond to 
the resolution of a singularity $H_1$ ($H_2$) 
at $z=0$ ($z=\infty$), except for 
$\nu_0$ and $\tx\nu_0$. The volumes of the generic elliptic fiber $E$ and
the base $B$ correspond to the linear relations
\eqn\applmlii{
l^{(E)}:\ -6e_0+2e_2+3e_3+f_1=0\ ,\qquad
l^{(B)}:\ v_0+\tx v_0-2f_1=0\ ,}
where $e_0=(0,0,0)$ is the single interior point of the polyhedron 
$\Delta_{M_2}$ and the coefficient of $e_0$ is determined by the fact that
the vertices $\nu_i$ are in fact a short-hand notation for 
four-dimensional vertices in the hyperplane $x_0=1$ of a 
four-dimensional integral lattice $\Lambda_4$ \TRV.

The linear relation $l^{(B)}$ determines a holomorphic curve $C_B$
which is isomorphic to the base of the elliptic fibration of $M_2$.
The local limit in K\"ahler moduli is thus given by $t_B\to i\infty$, where
$t_B$ is the flat coordinate at large radius whose imaginary part 
measures the size of the base $B$.

In the K3 manifold $M_2$, a vertex $\nu_i$ of $\Delta_{M_2}$ corresponds to a
divisor $D_i:\; x_i=0$ in $M_2$, where $x_i$ are the Batyrev-Cox variables
used in the definition of the hyperplane as in \batpol.
In particular, $C_B\subset M_2$
is given by $x_{f_1}=0$. 
There are further linear relations, or equivalently K\"ahler classes,
$l^{(+,r)}$, which 
are associated to the resolution of the singularities $H_1$. They 
involve only vertices in $N_0$ and $N_+$. Similarly there is another
set of linear relations $l^{(-,r)}$ involving vertices in 
$N_0\cup N_-$ associated to the resolution of the $H_-$ singularity.
In terms of the mirror manifold $W_2$, 
the vertices $\nu_i$ correspond to deformations of the defining 
equation in \batpol. In particular a vertex $\nu_i\subset N_\Sigma$
corresponds to a monomial in $p_\Sigma$ in \ktm. To find 
the limit in complex structure corresponding to $t_B\to i\infty$ we
need also the action of mirror symmetry on the moduli space, or said
differently the map between K\"ahler moduli of $M_2$
and complex structure moduli of $W_2$. In the large radius/large 
complex structure limit it is given in terms of the so-called 
algebraic coordinates $z_r$ by
\eqn\applmliv{
t_r=\ov{1}{2\pi i} \ln z_r,\hskip 2cm
z_r=\prod_i a_i^{l^{(r)}_i}\ ,}
where $a_i$ are the parameters for the complex structure related to 
the vertex $\nu_i$ as defined in \batpol.

We are now ready to establish the validity of \lml. We are searching for a
limit $a_i\to \eps^{\lambda_i}a_i$ of the parameters $a_i$ in \batpol, 
defined by 
the exponents $\lambda_i$, which corresponds
to $t_B\to i\infty$ for $\eps\to0$,
while keeping all the other K\"ahler moduli fixed. 
{}From \applmliv\ it follows that $\la_i=\sum_j c^j\nu_{i,j}$ 
leaves all $z_r$ invariant, with $\nu_{i,j}$
denoting the $j$-th entry of the vertex $\nu_i$ and $c^j$ constants. 
However there is another solution for the $\la_i$ that leaves 
invariant the $z_{\pm,r}$ as well as $z_E$, namely 
$a_{\Sigma,i}\to \eps^{\la_{\Sigma,i}}a_{\Sigma,i}$ with 
$\la_{\Sigma,i}=c_\Sigma\nu_{i,1}$ parametrized by three constants
$c_\Sigma$ of which $c_0$ is irrelevant. The only two classes of 
inequivalent solutions for the 
constants $c_+,c_-$ that are compatible 
with $z_B\to 0$, are $c^1_+=1,\ c^1_-=0$ 
or $c^1_+=0,\ c^1_-=-1$.
The first entry $\nu_{i,1}$
determines also the $v$ power of the monomial $\prod_j x_j^{1+\langle\nus_j,
\nu_i\rangle}$ multiplying $a_i$. So $z_{+,r}$ will be independent
of $\eps\to 0$ precisely if $\lambda_{+,i}$ is proportional to the 
$v$ power of the monomial that multiplies $a_{+,i}$. Therefore the two
solutions are precisely
the two patches of the local mirror limit in \lml, in agreement
with our assertion.

For the higher-dimensional case, we can similarly split the vertices $\nu_i$
according to the sign of the $n$-th entry $\nu_{i,n}$, which again 
determines the power of $v$ of the monomial associated to $\nu_i$ in
the mirror polynomial $p_{\Delta^*}$. The rest follows from the above
$n=1$ case.
 
\appendix{B}{Topological data for the manifolds in sect. 7}
The following tables specify the 
topological data for the elliptic fibrations above the bases $\tIF_k$ 
and $\tIF_{k,m,n}$ described in section 7. The hodge numbers
are given in the first line in the form $(h^{1,1}_{\delta h^{1,1}},
h^{1,2},h^{1,3}_{\delta h^{1,3}})$, where as before $\delta h^{1,i}$ 
denotes the number of $h^{1,i}$ deformations that are frozen
in the toric model. The second line denotes the Euler number
$\chi$ and $(\chi \; {\rm mod} \; 24)$.
$$
\vbox{\offinterlineskip\tabskip=0pt\halign{
\strut
\vrule\hfil~$\ss{#}$~\hfil&\hfil~$\ss{#}$~\hfil&\hfil~$\ss{#}$~\hfil&
\hfil~$\ss{#}$~\hfil&\hfil~$\ss{#}$~\hfil&\hfil~$\ss{#}$~\hfil\vrule\cr
\noalign{\hrule}
SU_1&SU_2&SU_3&SU_4&SU_5&SU_6\cr
\noalign{\hrule}
&&k=-18&G^{(2)}=SU_1&&\cr
(11_0,0,30989_0)&\qquad\qquad&\qquad\qquad&\qquad\qquad&\qquad\qquad&\qquad\qquad\cr 
      186048 \;  (0)&&&&&\cr
\noalign{\hrule}
&&k=-17&G^{(2)}=SU_1&&\cr
(12_0,0,28348_0)&&&&&\cr 
170208 \;  (0)&&&&&\cr
\noalign{\hrule}
&&k=-16&G^{(2)}=SU_1&&\cr
(12_0,0,25828_0)&&&&& \cr
 155088 \;  (0)&&&&&\cr 
\noalign{\hrule}
&&k=-15&G^{(2)}=SU_1&&\cr
(12_0,1,23429_0)&&&&&\cr 
140688 \;  (0)&&&&&\cr 
\noalign{\hrule}
&&k=-14&G^{(2)}=SU_1&&\cr
(12_0,3,21151_0)&&&&&\cr 
127008 \;  (0)&&&&&\cr
\noalign{\hrule}
&&k=-13&G^{(2)}=SU_1&&\cr
(12_0,6,18994_0)&&&&&\cr
114048 \;  (0)&&&&&\cr
\noalign{\hrule}
&&k=-12&G^{(2)}=SU_1&&\cr
(12_0,10,16958_0)& 
      (10_0,0,16959_0)&&&&\cr
101808 \;  (0)& 
      101862 \;  (6)&&&&\cr
\noalign{\hrule}
&&k=-11&G^{(2)}=SU_1&&\cr
(12_0,15,15043_0)& 
      (10_0,0,15046_0)&&&&\cr 
90288 \;  (0)& 
      90384 \;  (0)&&&&\cr
\noalign{\hrule}
&&k=-10&G^{(2)}=SU_1&&\cr
(12_0,21,13249_0)& 
      (10_0,0,13255_0)&&&&\cr
79488 \;  (0)& 
      79638 \;  (6)&&&&\cr
\noalign{\hrule}
}}
$$
$$
\vbox{\offinterlineskip\tabskip=0pt\halign{
\strut
\vrule\hfil~$\ss{#}$~\hfil&\hfil~$\ss{#}$~\hfil&\hfil~$\ss{#}$~\hfil&
\hfil~$\ss{#}$~\hfil&\hfil~$\ss{#}$~\hfil&\hfil~$\ss{#}$~\hfil\vrule\cr
\noalign{\hrule}
SU_1&SU_2&SU_3&SU_4&SU_5&SU_6\cr
\noalign{\hrule}
&&k=-9&G^{(2)}=SU_1&&\cr
(12_0,28,11576_0)& 
      (10_0,0,11586_0)& 
      (9_0,0,11587_0)&&&\cr
69408 \;  (0)& 
      69624 \;  (0)& 69624 \;  (0)&&&\cr
\noalign{\hrule}
&&k=-8&G^{(2)}=SU_1&&\cr
(12_0,36,10024_0)& 
      (10_0,0,10039_0)& 
      (9_0,0,10042_0)&&&\cr
60048 \;  (0)& 
      60342 \;  (6)& 60354 \;  (18)&&&\cr
\noalign{\hrule}
&&k=-7&G^{(2)}=SU_1&&\cr
(12_0,45,8593_0)& 
      (10_0,0,8614_0)& 
      (9_0,0,8620_0)&&&\cr
51408 \;  (0)& 
      51792 \;  (0)& 51822 \;  (6)&&&\cr
\noalign{\hrule}
&&k=-6&G^{(2)}=SU_1&&\cr
(12_0,55,7283_0)& 
      (10_0,0,7311_0)& 
      (9_0,0,7321_0)& 
      (8_0,0,7322_0)&&\cr
43488 \;  (0)& 
      43974 \;  (6)& 44028 \;  (12)& 
      44028 \;  (12)&&\cr
\noalign{\hrule}
&&k=-5&G^{(2)}=SU_1&&\cr
(12_0,66,6094_0)& 
      (10_0,0,6130_0)& 
      (9_0,0,6145_0)& 
      (8_0,0,6148_0)&&\cr
36288 \;  (0)& 
      36888 \;  (0)& 36972 \;  (12)& 
      36984 \;  (0)&&\cr
\noalign{\hrule}
&&k=-4&G^{(2)}=SU_1&&\cr
(12_0,78,5026_0)& 
      (10_0,0,5071_0)& 
      (9_0,0,5092_0)& 
      (8_0,0,5098_0)&&\cr
29808 \;  (0)& 
      30534 \;  (6)& 30654 \;  (6)& 
      30684 \;  (12)&&\cr
\noalign{\hrule}
&&k=-3&G^{(2)}=SU_1&&\cr
(12_0,91,4079_0)& 
      (10_0,0,4134_0)& 
      (9_0,0,4162_0)& 
      (8_0,0,4172_0)& 
      (7_0,0,4173_0)&\cr
24048 \;  (0)& 
      24912 \;  (0)& 25074 \;  (18)& 
      25128 \;  (0)& 25128 \;  (0)&\cr
\noalign{\hrule}
&&k=-2&G^{(2)}=SU_1&&\cr
(12_0,105,3253_0)& 
      (10_0,0,3319_0)& 
      (9_0,0,3355_0)& 
      (8_0,0,3370_0)& 
      (7_0,0,3373_0)&\cr
19008 \;  (0)& 
      20022 \;  (6)& 20232 \;  (0)& 
      20316 \;  (12)& 20328 \;  (0)&\cr
\noalign{\hrule}
&&k=-1&G^{(2)}=SU_1&&\cr
(12_0,120,2548_0)& 
      (10_0,0,2626_0)& 
      (9_0,0,2671_0)& 
      (8_0,0,2692_0)& 
      (7_0,0,2698_0)&\cr
14688 \;  (0)& 
      15864 \;  (0)& 16128 \;  (0)& 
      16248 \;  (0)& 16278 \;  (6)&\cr
\noalign{\hrule}
&&k=0&G^{(2)}=SU_1&&\cr
(12_0,136,1964_0)& 
      (10_0,0,2055_0)& 
      (9_0,0,2110_0)& 
      (8_0,0,2138_0)& 
      (7_0,0,2148_0)& 
      (6_0,0,2149_0)\cr
 11088 \;  (0)& 12438 \;  (6)& 
      12762 \;  (18)& 12924 \;  (12)& 
      12978 \;  (18)& 12978 \;  (18)\cr 
\noalign{\hrule} 
&&k=1&G^{(2)}=SU_1&&\cr
      (12_0,153,1501_0)& 
      (10_0,0,1606_0)& 
      (9_0,0,1672_0)& 
      (8_0,0,1708_0)& 
      (7_0,0,1723_0)& 
      (6_0,0,1726_0)\cr
      8208 \;  (0)& 9744 \;  (0)& 
      10134 \;  (6)& 10344 \;  (0)& 
      10428 \;  (12)& 10440 \;  (0)\cr 
\noalign{\hrule} 
}}
$$
$$
\vbox{\offinterlineskip\tabskip=0pt\halign{
\strut
\vrule\hfil~$\ss{#}$~\hfil&\hfil~$\ss{#}$~\hfil&\hfil~$\ss{#}$~\hfil&
\hfil~$\ss{#}$~\hfil&\hfil~$\ss{#}$~\hfil&\hfil~$\ss{#}$~\hfil\vrule\cr
\noalign{\hrule}
SU_1&SU_2&SU_3&SU_4&SU_5&SU_6\cr
\noalign{\hrule}
&&k=2&G^{(2)}=SU_1&&\cr
(12_0,171,1159_0)& 
      (10_0,0,1279_0)& 
      (9_0,0,1357_0)& 
      (8_0,0,1402_0)& 
      (7_0,0,1423_0)& 
      (6_0,0,1429_0)\cr 
      6048 \;  (0)& 7782 \;  (6)& 
      8244 \;  (12)& 8508 \;  (12)& 
      8628 \;  (12)& 8658 \;  (18)\cr\noalign{\hrule}
&&k=3&G^{(2)}=SU_1&&\cr
  (12_0,191,939_0)& 
      (10_0,1,1075_0)& 
      (9_0,1,1166_0)& 
      (8_0,1,1221_0)& 
      (7_0,1,1249_0)& 
      (6_0,1,1259_0)\cr 
      4608 \;  (0)& 6552 \;  (0)& 
      7092 \;  (12)& 7416 \;  (0)& 
      7578 \;  (18)& 7632 \;  (0)\cr \noalign{\hrule} 
&&k=4&G^{(2)}=SU_2&&\cr
      (13_0,210,828_0)& 
      (11_0,0,981_0)& 
      (10_0,0,1086_0)& 
      (9_0,0,1152_0)& 
      (8_0,0,1188_0)& 
      (7_0,0,1203_0)\cr 
      3834 \;  (18)& 6000 \;  (0)& 
      6624 \;  (0)& 7014 \;  (6)& 
      7224 \;  (0)& 7308 \;  (12)\cr \noalign{\hrule} 
&&k=5&G^{(2)}=G_2&&\cr
      (14_0,231,773_0)& 
      (12_0,0,944_0)& 
      (11_0,0,1064_0)& 
      (10_0,0,1142_0)& 
      (9_0,0,1187_0)& 
      (8_0,0,1208_0)\cr
      3384 \;  (0)& 5784 \;  (0)& 
      6498 \;  (18)& 6960 \;  (0)& 
      7224 \;  (0)& 7344 \;  (0)\cr \noalign{\hrule} 
&&k=6&G^{(2)}=G_2&&\cr
      (16_2,253,745_0)& 
      (14_2,0,935_0)& 
      (13_2,0,1071_0)& 
      (12_2,0,1162_0)& 
      (11_2,0,1217_0)& 
      (10_2,0,1245_0)\cr
      3096 \;  (0)& 5742 \;  (6)& 
      6552 \;  (0)& 7092 \;  (12)& 
      7416 \;  (0)& 7578 \;  (18)\cr \noalign{\hrule} 
&&k=7&G^{(2)}=F_4&&\cr
     (16_0,276,736_0)& 
      (14_0,0,946_0)& 
      (13_0,0,1099_0)& 
      (12_0,0,1204_0)& 
      (11_0,0,1270_0)& 
      (10_0,0,1306_0)\cr 
      2904 \;  (0)& 5808 \;  (0)& 
      6720 \;  (0)& 7344 \;  (0)& 
      7734 \;  (6)& 7944 \;  (0)\cr 
\noalign{\hrule}
&&k=8&G^{(2)}=F_4&&\cr
(16_0,300,736_0)& 
      (14_0,0,967_0)& 
      (13_0,0,1138_0)& 
      (12_0,0,1258_0)& 
      (11_0,0,1336_0)& 
      (10_0,0,1381_0)\cr 
      2760 \;  (0)& 5934 \;  (6)& 
      6954 \;  (18)& 7668 \;  (12)& 
      8130 \;  (18)& 8394 \;  (18)\cr 
\noalign{\hrule}
&&k=9&G^{(2)}=F_4&&\cr
(18_2,325,743_0)& 
      (16_2,0,996_0)& 
      (15_2,0,1186_0)& 
      (14_2,0,1322_0)& 
      (13_2,0,1413_0)& 
      (12_2,0,1468_0)\cr 
      2664 \;  (0)& 6120 \;  (0)& 
      7254 \;  (6)& 8064 \;  (0)& 
      8604 \;  (12)& 8928 \;  (0)\cr
\noalign{\hrule}
&&k=10&G^{(2)}=E_7&&\cr
(19_0,351,757_0)& 
      (17_0,0,1033_0)& 
      (16_0,0,1243_0)& 
      (15_0,0,1396_0)& 
      (14_0,0,1501_0)& 
      (13_0,0,1567_0)\cr 
      2598 \;  (6)& 6348 \;  (12)& 
      7602 \;  (18)& 8514 \;  (18)& 
      9138 \;  (18)& 9528 \;  (0)\cr
\noalign{\hrule}
&&k=11&G^{(2)}=E_7&&\cr
(19_0,378,775_0)& 
      (17_0,0,1075_0)& 
      (16_0,0,1306_0)& 
      (15_0,0,1477_0)& 
      (14_0,0,1597_0)& 
      (13_0,0,1675_0)\cr 
      2544 \;  (0)& 6600 \;  (0)& 
      7980 \;  (12)& 9000 \;  (0)& 
      9714 \;  (18)& 10176 \;  (0)\cr
\noalign{\hrule}
&&k=12&G^{(2)}=E_7&&\cr
(19_0,406,796_0)& 
      (17_0,0,1121_0)& 
      (16_0,0,1374_0)& 
      (15_0,0,1564_0)& 
      (14_0,0,1700_0)& 
      (13_0,0,1791_0)\cr 
      2502 \;  (6)& 6876 \;  (12)& 
      8388 \;  (12)& 9522 \;  (18)& 
      10332 \;  (12)& 10872 \;  (0)\cr
\noalign{\hrule}
}}
$$
$$
\vbox{\offinterlineskip\tabskip=0pt\halign{
\strut
\vrule\hfil~$\ss{#}$~\hfil&\hfil~$\ss{#}$~\hfil&\hfil~$\ss{#}$~\hfil&
\hfil~$\ss{#}$~\hfil&\hfil~$\ss{#}$~\hfil&\hfil~$\ss{#}$~\hfil\vrule\cr
\noalign{\hrule}
SU_1&SU_2&SU_3&SU_4&SU_5&SU_6\cr
\noalign{\hrule}
&&k=13&G^{(2)}=E_8\times\tilde{SU}_2&&\cr
(21_0,441,820_0)& 
      (19_0,6,1171_0)& 
      (18_0,6,1447_0)& 
      (17_0,6,1657_0)& 
      (16_0,6,1810_0)& 
      (15_0,6,1915_0)\cr 
      2448 \;  (0)& 7152 \;  (0)& 
      8802 \;  (18)& 10056 \;  (0)& 
      10968 \;  (0)& 11592 \;  (0)\cr
\noalign{\hrule}
&&k=14&G^{(2)}=E_8\times\tilde{SU}_2&&\cr
(21_0,468,847_0)& 
      (19_0,3,1225_0)& 
      (18_0,3,1525_0)& 
      (17_0,3,1756_0)& 
      (16_0,3,1927_0)& 
      (15_0,3,2047_0)\cr 
      2448 \;  (0)& 7494 \;  (6)& 
      9288 \;  (0)& 10668 \;  (12)& 
      11688 \;  (0)& 12402 \;  (18)\cr
\noalign{\hrule}
&&k=15&G^{(2)}=E_8\times\tilde{SU}_2&&\cr
(21_0,497,876_0)& 
      (19_0,1,1282_0)& 
      (18_0,1,1607_0)& 
      (17_0,1,1860_0)& 
      (16_0,1,2050_0)& 
      (15_0,1,2186_0)\cr 
      2448 \;  (0)& 7848 \;  (0)& 
      9792 \;  (0)& 11304 \;  (0)& 
      12438 \;  (6)& 13248 \;  (0)\cr
\noalign{\hrule}
&&k=16&G^{(2)}=E_8\times\tilde{SU}_2&&\cr
(21_0,528,907_0)& 
      (19_0,0,1342_0)& 
      (18_0,0,1693_0)& 
      (17_0,0,1969_0)& 
      (16_0,0,2179_0)& 
      (15_0,0,2332_0)\cr 
      2448 \;  (0)& 8214 \;  (6)& 
      10314 \;  (18)& 11964 \;  (12)& 
      13218 \;  (18)& 14130 \;  (18)\cr
\noalign{\hrule}&&k=17&G^{(2)}=E_8\times\tilde{SU}_2&&\cr
(21_0,561,940_0)& 
      (19_0,0,1405_0)& 
      (18_0,0,1783_0)& 
      (17_0,0,2083_0)& 
      (16_0,0,2314_0)& 
      (15_0,0,2485_0)\cr 
      2448 \;  (0)& 8592 \;  (0)& 
      10854 \;  (6)& 12648 \;  (0)& 
      14028 \;  (12)& 15048 \;  (0)\cr
\noalign{\hrule}
&&k=18&G^{(2)}=E_8&&\cr
(20_0,595,975_0)& 
      (18_0,0,1471_0)& 
      (17_0,0,1877_0)& 
      (16_0,0,2202_0)& 
      (15_0,0,2455_0)& 
      (14_0,0,2645_0)\cr 
      2448 \;  (0)& 8982 \;  (6)& 
      11412 \;  (12)& 13356 \;  (12)& 
      14868 \;  (12)& 16002 \;  (18)\cr
\noalign{\hrule}}}
$$
\leftskip .5cm \rightskip .5cm 
\noindent{\ninepoint  \baselineskip=8pt  
{{\bf Table B.1:} Topological data for the elliptic fibrations
over $\tIF_{k}$ dual to $SU(N)$ bundles on elliptic fibrations
$Z_3\to \IP^2$ as described in sect. 7.2.}}

\leftskip .0cm \rightskip .0cm \ni

$$
\vbox{\offinterlineskip\tabskip=0pt\halign{
\strut
\vrule\hfil~$\ss{#}$~\hfil&\hfil~$\ss{#}$~\hfil&\hfil~$\ss{#}$~\hfil&
\hfil~$\ss{#}$~\hfil&\hfil~$\ss{#}$~\hfil&\hfil~$\ss{#}$~\hfil\vrule\cr
\noalign{\hrule}
n=7&n=8&n=9&n=10&n=11&n=12\cr
(51_0,763,1252_0)&  (51_0,827,1352_0)&    (72_0,886,1454_0)& 
 (72_0,950,1558_0)&   (72_0,1014,1662_0)&  (73_1,1079,1766_0)\cr 
      3288 \;  (0)& 
      3504 \;  (0)& 
      3888 \;  (0)& 
      4128 \;  (0)& 
      4368 \;  (0)& 
      4608 \;  (0)\cr
\noalign{\hrule}}}
$$
\leftskip .5cm \rightskip .5cm 
\noindent{\ninepoint  \baselineskip=8pt  
{{\bf Table B.2:} Topological data for the elliptic fibrations
over $\tIF_{12,12-6n,n}$ dual to $SU(1)$ bundles on
elliptic fibrations $Z_3\to \IFO_n$ as described in sect. 7.3.}}

\leftskip .0cm \rightskip .0cm \ni
$$
\vbox{\offinterlineskip\tabskip=0pt\halign{
\strut
\vrule\hfil~$\ss{#}$~\hfil&\hfil~$\ss{#}$~\hfil&\hfil~$\ss{#}$~\hfil&
\hfil~$\ss{#}$~\hfil&\hfil~$\ss{#}$~\hfil&\hfil~$\ss{#}$~\hfil\vrule\cr
\noalign{\hrule}
SU_1&SU_2&SU_3&SU_4&SU_5&SU_6\cr
\noalign{\hrule}
&&n=-6&G^{(2)}=E_8&&\cr
(41_0,707,1154_0)& 
      (39_0,2,1739_0)& 
      (38_0,2,2212_0)& 
      (37_0,2,2581_0)& 
      (36_0,2,2854_0)& 
      (35_0,2,3039_0)\cr 
      2976 \;  (0)& 10704 \;  (0)& 
      13536 \;  (0)& 15744 \;  (0)& 
      17376 \;  (0)& 18480 \;  (0)\cr 
\noalign{\hrule}
&&n=-5&G^{(2)}=E_8&&\cr
(41_0,645,1060_0)& 
      (39_0,0,1593_0)& 
      (38_0,0,2022_0)& 
      (37_0,0,2355_0)& 
      (36_0,0,2600_0)& 
      (35_0,0,2765_0)\cr 
      2784 \;  (0)& 9840 \;  (0)& 
      12408 \;  (0)& 14400 \;  (0)& 
      15864 \;  (0)& 16848 \;  (0)\cr
\noalign{\hrule}
&&n=-4&G^{(2)}=E_8&&\cr
(31_0,597,974_0)& 
      (28_0,2,1470_0)& 
      (28_0,2,1872_0)& 
      (27_0,2,2191_0)& 
      (26_0,2,2434_0)& 
      (25_0,2,2609_0)\cr 
      2496 \;  (0)& 9024 \;  (0)& 
      11436 \;  (12)& 13344 \;  (0)& 
      14796 \;  (12)& 15840 \;  (0)\cr
\noalign{\hrule}
&&n=-3&G^{(2)}=E_8&&\cr
(26_0,552,902_0)& 
      (24_0,1,1361_0)& 
      (23_0,1,1736_0)& 
      (22_0,1,2035_0)& 
      (21_0,1,2266_0)& 
      (20_0,1,2437_0)\cr 
      2304 \;  (0)& 8352 \;  (0)& 
      10596 \;  (12)& 12384 \;  (0)& 
      13764 \;  (12)& 14784 \;  (0)\cr
\noalign{\hrule}
&&n=-2&G^{(2)}=E_8&&\cr
(21_0,529,868_1)& 
      (19_0,0,1309_1)& 
      (18_0,0,1670_1)& 
      (17_0,0,1959_1)& 
      (16_0,0,2184_1)& 
      (15_0,0,2353_1)\cr 
      2208 \;  (0)& 8016 \;  (0)& 
      10176 \;  (0)& 11904 \;  (0)& 
      13248 \;  (0)& 14256 \;  (0)\cr
\noalign{\hrule}
&&n=-1&G^{(2)}=E_8&&\cr
(21_0,529,868_0)& 
      (19_0,0,1309_0)& 
      (18_0,0,1670_0)& 
      (17_0,0,1959_0)& 
      (16_0,0,2184_0)& 
      (15_0,0,2353_0)\cr
      2208 \;  (0)& 8016 \;  (0)& 
      10176 \;  (0)& 11904 \;  (0)& 
      13248 \;  (0)& 14256 \;  (0)\cr
\noalign{\hrule}
&&n=0&G^{(2)}=E_8&&\cr
(21_0,529,868_0)& 
      (19_0,0,1309_0)& 
      (18_0,0,1670_0)& 
      (17_0,0,1959_0)& 
      (16_0,0,2184_0)& 
      (15_0,0,2353_0)\cr 
      2208 \;  (0)& 8016 \;  (0)& 
      10176 \;  (0)& 11904 \;  (0)& 
      13248 \;  (0)& 14256 \;  (0)\cr 
\noalign{\hrule}}}
$$
\leftskip .5cm \rightskip .5cm 
\noindent{\ninepoint  \baselineskip=8pt  
{{\bf Table B.3:} Topological data for the elliptic fibrations
over $\tIF_{12,12-6n,n}$ dual to $SU(N)$ bundles on
elliptic fibrations $Z_3\to \IFO_n$ as described in sect. 7.3.}}
\leftskip .0cm \rightskip .0cm \ni

\listrefs
\end